\newif\ifnatbibstyle
\natbibstylefalse
\newif\ifnotnatbibstyle
\ifnatbibstyle\notnatbibstylefalse\else\notnatbibstyletrue\fi

\ifnatbibstyle
\PassOptionsToPackage{sort}{natbib}
\fi
\documentclass[10pt,twocolumn]{sigplanconf}
\usepackage{epsfig}
\usepackage{hyperref}
\usepackage{amsfonts, amssymb, amsmath, fancyhdr, color, url, amsthm}
\usepackage{graphicx}
\usepackage{dsfont}
\usepackage{datetime}
\usepackage{authblk}
\usepackage{array}
\usepackage{enumitem}
\usepackage{comment}
\usepackage{alltt}
\usepackage{listings}
\usepackage{xfrac}
\usepackage{url}
\usepackage{mathptmx}
\newcommand{\ignore}[1]{}

\newif\iftechreport
\newif\ifnontechreport
\techreporttrue
\iftechreport\nontechreportfalse\else\nontechreporttrue\fi
\newcommand{\techreport}[1]{\iftechreport #1 \fi}
\newcommand{\nontechreport}[1]{\ifnontechreport #1 \fi}
\let\oldcaption\caption
\renewcommand{\caption}[1]{\oldcaption{\normalsize #1}}

\ifnotnatbibstyle

\let\bibfont\relax

\expandafter\let\csname ver@natbib.sty\endcsname\relax
\techreport{\usepackage[maxnames=20]{biblatex}}
\nontechreport{\usepackage[maxnames=2]{biblatex}}
\fi

\definecolor{dkgreen}{rgb}{0,0.6,0}
\definecolor{gray}{rgb}{0.5,0.5,0.5}
\definecolor{mauve}{rgb}{0.58,0,0.82}
\newcommand{\myvspace}[1]{\vspace{#1}}
\techreport{\renewcommand{\myvspace}[1]{}}

\usepackage{wasysym}
\usepackage{subfigure}
\usepackage{xspace}

\newcommand{\fixthis}[1]{\textcolor{red}{\textbf{FIX THIS: #1}}}
\newcommand{\remark}[1]{\textcolor{blue}{\textbf{[#1]}}}

\makeatletter
\usepackage[compact]{titlesec}
\let\subparagraph\relax 

\titlespacing*\section{0pt}{6pt plus 4pt minus 2pt}{2pt plus 2pt minus 2pt}
\titlespacing*\subsection{0pt}{6pt plus 4pt minus 2pt}{2pt plus 2pt minus 2pt}

\makeatletter
\edef\@listi{\@listi \topsep2pt \parsep0pt \itemsep2pt}
\let\@listI=\@listi
\makeatother
\makeatother

\newcommand{\red}[1]{\begingroup \color{red} #1\endgroup}
\newcommand{\blue}[1]{\begingroup \color{blue} #1\endgroup}
\renewcommand{\blue}[1]{\begingroup #1\endgroup}
\newcommand{\highlight}[1]{\begingroup \color{blue} #1\endgroup}
\renewcommand{\highlight}[1]{\begingroup #1\endgroup}

\newcommand{\sys}{EyeQ++\xspace}

\renewcommand{\sys}{Parley\xspace}

  \makeatletter
  \let\@copyrightspace\relax
  \makeatother

\begin{document}
\sloppy
\title{Flexible Network Bandwidth and Latency Provisioning in the Datacenter}
\authorinfo{Vimalkumar Jeyakumar$^1$, Abdul Kabbani$^2$, Jeffrey C. Mogul$^2$, Amin Vahdat$^{2,3}$}
{$^1$Stanford University, $^2$Google, $^3$UCSD\vspace{-2.5em}}{}
\maketitle

\hyphenation{data-center}
\subsection*{Abstract}
Predictably sharing the network is critical to achieving high
utilization in the datacenter.  Past work has focussed on providing
bandwidth to endpoints, but often we want to allocate resources
among multi-node \emph{services}.
In this paper, we present \sys,
which provides service-centric minimum bandwidth guarantees,
which can be composed hierarchically.  \sys also supports
service-centric
weighted sharing of bandwidth in excess of these guarantees.
Further, we show how to configure these policies so services can get
low latencies even at high network load.  We evaluate \sys on a
multi-tiered oversubscribed network connecting 90 machines, each with
a 10Gb/s network interface, and demonstrate that \sys is able to meet its goals.


\section{Introduction}\label{sec:intro}
Multi-tenancy is inevitable at large scale.  \blue{In Google's
datacenters}, we see on average tens of jobs per server.  A typical
server is shared by applications that have diverse performance
requirements from the network, such as:

\begin{itemize}[noitemsep,leftmargin=1em,nolistsep]
  \item {\bf Bandwidth Intensive}: A MapReduce job during its read and
    shuffle stages; or, large file copies.
  \item {\bf Latency Sensitive}: The front-end for user-facing
    services, such as web search.
  \item {\bf Bandwidth and Latency Sensitive}: Infrastructure services,
    such a distributed file system (DFS), often have a mix of bandwidth
    and latency sensitive flows.
\end{itemize}
These applications can also be {\bf adversarial}.
For example, tenant virtual machines (VMs) sharing the same network,
    in public clouds such as Amazon EC2, Google Compute Engine,
    HP Cloud, and Windows Azure, are not likely to be cooperative.

Prior work~(see~\cite{mogul2012we} for a detailed summary) in this
space has already made the case for sharing bandwidth across entities
that are coarser-grained than a TCP flow.  For example,
Oktopus~\cite{ballani2011towards},
Gatekeeper~\cite{rodrigues2011gatekeeper},
EyeQ~\cite{jeyakumar2013eyeq} and
ElasticSwitch~\cite{popa2013elasticswitch} all provide, for a tenant's
collection of VMs, the abstraction of a dedicated physical network,
with a specified guaranteed bandwidth for each endpoint VM.  This
abstraction of a shared network can be useful for providers wishing to
support predictable behavior for tenant applications provisioned at a
VM granularity.

However, these prior systems that support bandwidth guarantees raise
the following questions:

\begin{itemize}[noitemsep,leftmargin=1em,nolistsep]
  \item How should we realize notions of sharing that are (a) more
    flexible than guarantees (e.g. \emph{weighted} sharing); and (b)
    service-centric, where services are a collection of endpoints?
  \item How should the service provision bandwidth to achieve
    service-level objectives, such as a bounded 99th percentile (tail)
    latency?
\end{itemize}

In this paper, we present \sys, a system we built to understand
answers to the above questions.  \sys has several features which make
it attractive for practical deployment: (1) it supports a mixture of
policies, such as bandwidth guarantees to service end-points, and
hierarchical, weighted sharing, constructed by nesting service
endpoints; 
(2) it uses a simple model of traffic characteristics to predict a
service's tail latency based on its own, and collocated services'
provisioned bandwidth, that is accurate at high loads.

We built \sys by systematically leveraging prior bandwidth sharing
systems to achieve our goals (we modify the open-sourced system EyeQ~\cite{jeyakumar2013eyeq}).
There are two parts to \sys: (a) static provisioning, which the
provider specifies when services are instantiated, and (b) runtime
provisioning, which \sys determines based on bandwidth usage.
\blue{While bandwidth guarantees are easy to statically provision
using admission control, \sys differs from EyeQ in that runtime
provisioning for hierarchical and weighted sharing requires
information about services' bandwidth usage.}

\begin{figure}[t]
  \centering\includegraphics[width=0.4\textwidth]{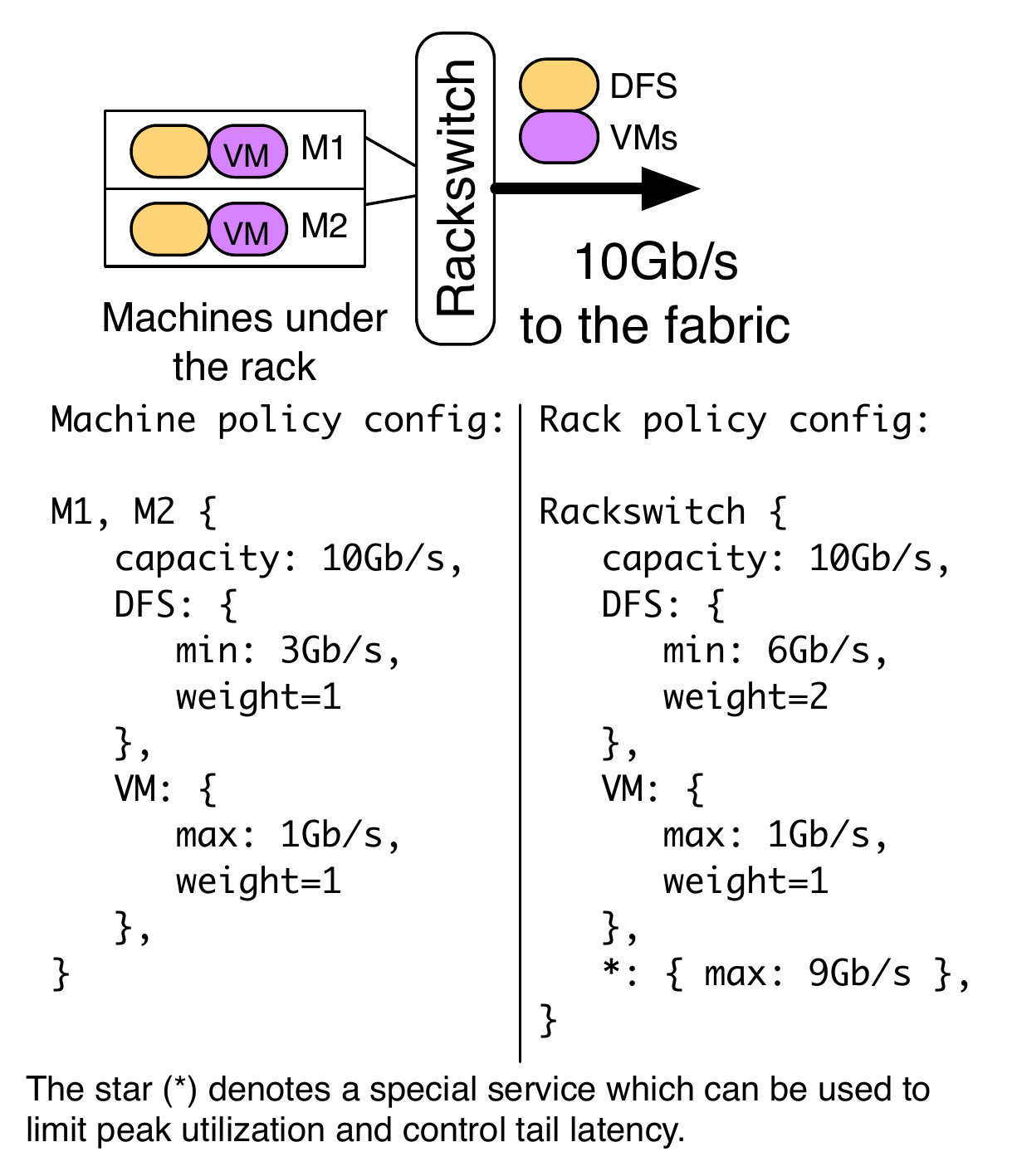}
  \caption{{\sys supports a hierarchical sharing policy.  The top
    figure shows the service placement on machines under the rack of a
    given capacity, and the bottom half illustrates sharing policies
    at the machine and rack.}}\label{fig:example}\myvspace{-2pt}
\end{figure}

For example, Figure~\ref{fig:example} shows an example of a static sharing policy,
in which the DFS service is allocated at least 6Gb/s, and at most
8Gb/s; and all (non-paying) VMs are allocated a maximum of 1Gb/s.
That aggregate of 1Gb/s is then shared in a `weighted max-min'~\cite[Sec.~6.5.2]{bertsekas1992data} fashion
among VMs in the rack.  \blue{Each VM's allocation depends on their own,
and the total bandwidth consumption of other VMs in the rack.  This
requires global service-level visibility of bandwidth usage
that EyeQ does not have.}

\ignore{
\begin{itemize}
\item the systematic
design of a hierarchical, service-oriented bandwidth-sharing
policy framework (\S\ref{sec:design}).
\item the use of \emph{optimization decomposition}~
to split a complex
problem into smaller, more tractable sub-problems (\S\ref{sec:decomp}).
\end{itemize}}


\blue{\sys adds this missing piece of global visibility on top of EyeQ,
while retaining its strengths, as follows:}
At the core of \sys's runtime system is a ``bandwidth broker'' that
allocates capacities to services in a hierarchical fashion.  At the
lowest level, each VM (or job, or service endpoint) on a machine is allocated an
aggregate transmit and receive \emph{hose capacity}.  These capacities are
periodically tuned by the
bandwidth broker \blue{to conform to the global policy},
based on (i) service endpoint's machine-level utilization
(\S\ref{subsec:machineshaper}), (ii) service's rackswitch-level utilization
(\S\ref{subsec:rackbroker}), and (iii) service's overall fabric utilization
(\S\ref{subsec:fabricbroker}).  Hose capacities are enforced directly
in the \emph{dataplane} using distributed, end-to-end congestion control.  This
decomposition of the global sharing objective into rack and machine
sub-problems enables \sys to scale to a large datacenter.

Specifically, the main contributions of our paper are:
\begin{itemize}
  \item The design and implementation of \sys that manages datacenter
    network bandwidth among services, in a flexible, and scalable
    fashion.  \highlight{\sys adds support for hierarchical allocations
    and distributed rate limiting, generalizing the notion of
    bandwidth guarantees provided by systems such as EyeQ and ElasticSwitch.}
  \item Demonstrating that a service's soft-realtime latency
    requirements can be met by appropriately controlling peak network
    load.
\end{itemize}
\ignore{
The remainder of the paper discusses the system in more detail.  We
begin with an overview of requirements for predictable network
sharing, and then delve into the design of \sys in \S\ref{sec:design}.
In \S\ref{sec:latency}, we show how to provision bandwidth in light of
a service's latency requirements.  We then conclude with the
implementation~(\S\ref{sec:implementation}) and
evaluation~(\S\ref{sec:evaluation}) of \sys.
}
\ignore{
In \S\ref{sec:design}, we explicitly describe our sharing objective,
and we describe our algorithm to compute bandwidth shares.  In
\S\ref{sec:latency}, we show that these guarantees are useful
predictors of tail latency.  In \S\ref{sec:evaluation}, we evaluate
\sys's runtime in a real cluster of 90 machines, and show that that
our implementation (\S\ref{sec:implementation}) can maintain both
bandwidth guarantees and latency protection, while using minimal
computational resources.
}

\section{Background and Requirements}
\label{sec:background}
\begin{figure}[t]
  \centering\includegraphics[width=0.4\textwidth]{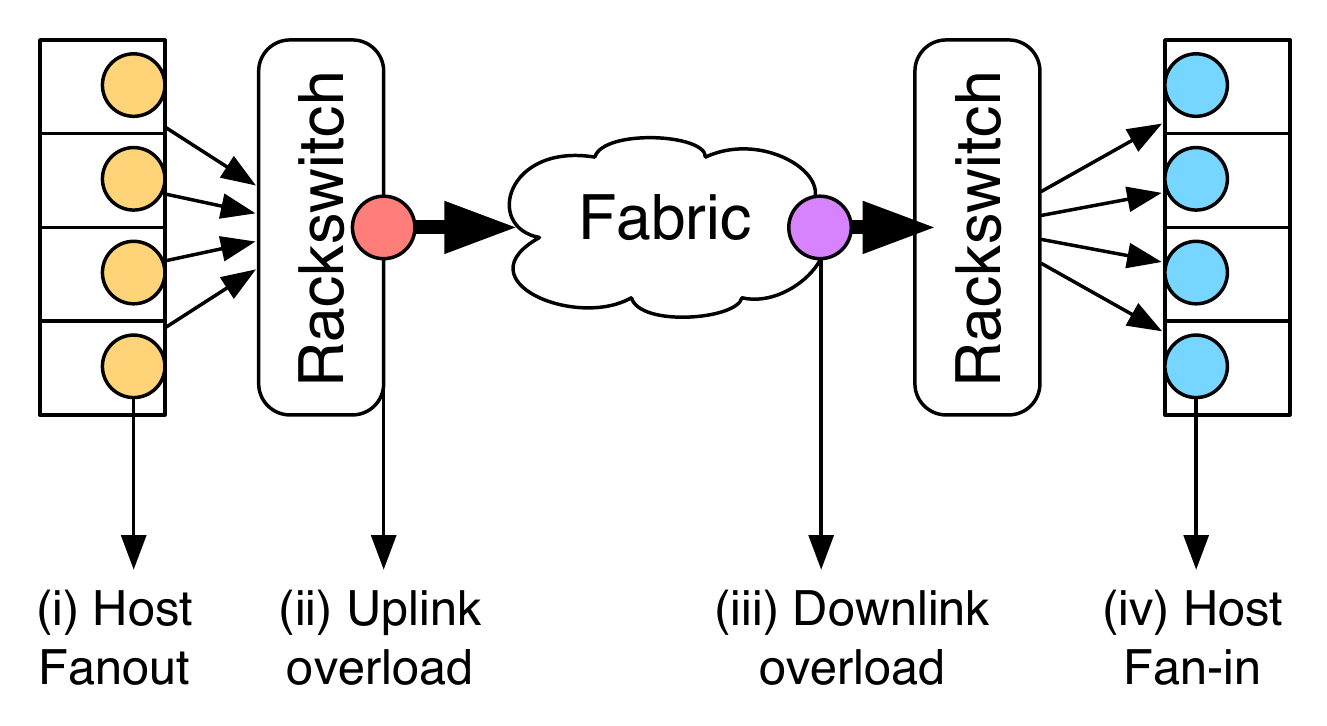}
  \caption{The contention points for intra-cluster bandwidth that \sys
    addresses.}\label{fig:contention}
\end{figure}

Datacenter bandwidth is not free.
Modern datacenter network topologies are typically multi-staged Clos
topologies~\cite{greenberg2009vl2} with few
over-subscription points (typically only at the top-of-rack switch).
Measurements from a highly-utilized datacenter 
show that packet drops occur predominantly at such over-subscription
points in the topology, as shown in Figure~\ref{fig:contention}.  The
contention can be broadly classified into three cases:


\begin{itemize}[noitemsep,leftmargin=1em,nolistsep]
  \item {\bf Host level}: multiple jobs sharing a host share the host (NIC)
    transmit and receive bandwidth.  On the transmit side, the host
    network stack is back-pressured, so we do not see any drops.
    However, a large fraction of dropped packets happen at the
    receiver, due to incast-like traffic patterns.  We call this
    \emph{host fanin}.
  \item {\bf First rackswitch}: Due to rackswitch over-subscription,
    hosts under a rack can overwhelm the limited-capacity uplinks.  We
    call this the \emph{uplink overload}.
  \item {\bf Last rackswitch}: Finally, some small fraction of drops
    happen \emph{within the fabric}, because the fabric overwhelms the
    rackswitch's limited capacity to its hosts.  We call this
    \emph{downlink overload}.
\end{itemize}

Very few drops occur elsewhere within the fabric.  We observe similar
trends in other highly-utilized clusters.

Our goal is to address contention-induced packet drops, because TCP is
the dominant protocol in the datacenter, and its default response to
dropped packets aims at fair per-flow bandwidth allocation---without
taking service-level requirements into consideration.

\subsection{What does `predictability' mean?}
Our primarily goal is to share bandwidth in a
programmable and well-specified fashion, and to have low, bounded tail latency.
We argue that many seemingly different notions of predictability can
be mapped to offering guarantees on aggregate bandwidth at a
contention point.

{\bf Service-level guarantees.}  Consider large-scale data-analysis jobs
with multiple MapReduce stages,
which care most about total completion time.  The total job completion
time is a complex function of the number of CPUs (number of workers),
the available network bandwidth, and the number of barriers (stages)
in the job~\cite{ferguson2012jockey}.  Job completion time generally
does not depend on short-term network latency.
\sys allows the job scheduler to request bandwidth while accounting
for contention at the machine and rack level.  This explicit guarantee
allows the job scheduler to make better
decisions~\cite{ferguson2012jockey}
when placing jobs and orchestrating transfers, \emph{enabling}
predictable job completion times.

{\bf Latency.}  Infrastructure services also have have stringent
requirements on tail latency (e.g., $<$20ms response time for 1MB~IO
operations at the 99th percentile).  Achieving tail latency guarantees
is hard, as this fundamentally depends on the stochastic nature of
flow arrivals.  This fundamental dependence cannot be avoided.
However, given a model of flow arrivals, a bandwidth guarantee can
\emph{bound} tail completion times.

For example, if the inter-flow arrival times are exponentially
distributed, and their processing times (i.e. flow sizes) are also
exponentially distributed, then the distribution of flow completion
times can be explicitly written down as a function of the service
instance's allocated bandwidth.  That is, in a steady state
$M/M/1/{\rm FIFO}$ queue, the flow completion time $t$ (i.e. waiting
time + service time) has the probability density function~\cite{queueing}
\[
f(t)=
\begin{cases}
\mu(1-\rho)\exp\left({-\mu(1-\rho)t}\right) & t>0\\
0 & \text{otherwise.}
\end{cases}
\]

\noindent where $\mu$ is the average service \emph{rate} (a function
of the available capacity), and $\rho\in(0,1)$ is the average load on
queue.  For example, if the average flow size 1MB, $\mu$ is 1.25 per
ms at 10Gb/s; we can then deduce that 99\% of flow completion times
are $<$18.4ms.



By empirically measuring the distribution of flow sizes, service
times, and inter-arrival times (all of which change relatively
infrequently), we can do an offline simulation to derive a bandwidth
guarantee requirement that will support robust bounds on a flow's tail
latency.  Though the formula makes an assumption about the
distribution of flow sizes and their inter-arrival times, we discuss
in \S\ref{sec:latency} how we can bound latency without modeling or
making assumptions about the arrival process, using a simple counting
argument.

Thus, to deal with latency requirements, \sys can explicitly guarantee
aggregate capacity to service endpoints, and control the peak load
on the network.

\ignore{
\remark{I would drop this paragraph; I don't think it adds anything.}
{\bf Bandwidth caps}: For certain applications, there is an economic
incentive to exactly allocate what the application has paid for, and
nothing more.  Moreover, an operator might want to limit the total
bandwidth consumed by `scavenger class' of applications such as
free-tier VMs, regardless of available bandwidth.  In these scenarios, we
do not strive for work-conserving allocations.  The rationale (albeit
non-technical) is that network operators do not want users of the
cluster to be accustomed to better performance during periods of low
utilization, and expect the same at high utilization.}

{\bf Summary}: The main takeaway is that for many use cases,
explicit bandwidth guarantees at the level of service
endpoints (e.g. VMs), and control of the peak load on the network,
helps application-specific schedulers to maintain predictability
for their own metrics.  Hence, \sys provides knobs to share network bandwidth
in a flexible fashion, and strives to meet these requirements as
quickly as possible.

\subsection{Related Work}
\label{sec:Related Work}
Many works address bandwidth management within a
cluster.  Since we build on top of EyeQ~\cite{jeyakumar2013eyeq}, we
inherit its benefits over prior work such as
Seawall~\cite{shieh2011sharing},
Gatekeeper~\cite{rodrigues2011gatekeeper},
Oktopus~\cite{ballani2011towards}, and
FairCloud~\cite{popa2012faircloud}.  We therefore focus on more recent
work.

Both EyeQ~\cite{jeyakumar2013eyeq} and
ElasticSwitch~\cite{popa2013elasticswitch} only offer work-conserving
(and non work-conserving) bandwidth guarantees to individual service
endpoints.  CloudMirror~\cite{lee2013cloudmirror} argues for
asymmetric transmit and receive guarantees.  However, they are limited
in their ability to handle network congestion in a flexible fashion.
If the network is congested, bandwidth is shared in a max-min fashion
either across receiving endpoints (EyeQ), or across contending
source/destination host pairs (ElasticSwitch), which need not
necessarily match service-specific goals.
\highlight{In \S\ref{subsec:macro} we show how this flexibility helps
in controlling tail latency at high loads.}


The bandwidth enforcer in Google's B4 SDN WAN~\cite{jain2013b4}
allocates bandwidth at the granularity of
(QoS,src-datacenter,dst-datacenter) tunnels.  This approach does not
scale to allocations within a datacenter, due to the large number of
(source,destination) pairs.  Instead, \sys's optimization
decomposition decouples mechanisms for pairwise allocations between
source/destination hosts from service-level allocations, which helps
scalability.

Hadrian~\cite{ballani2013chatty} describes a hierarchical,
work-conserving hose model to guarantee a tenant a minimum
inter-tenant bandwidth.  However, Hadrian requires switch changes to
count the number of flows on a per-tenant basis, which changes quite
frequently.  It is unclear if this is practical, and whether Hadrian's
allocations are stable over time.  However, the policy framework we
describe in this work is rich enough to capture inter-tenant isolation
as well.

DRL~\cite{raghavan2007cloud} focuses on distributed rate limits by
having service endpoints exchange demands using a gossip-protocol.
However, distributed rate limits alone is insufficient in sharing bandwidth in a predictable fashion, as a service
can target its entire bandwidth share at a single machine.  Our
approach is inspired by DRL, but uses a hierarchical decomposition to
explicitly handle the major oversubscription points.  For instance, by aggregating service
endpoints at a machine and rack level, we flexibly share limited machine and rack
bandwidth.

Commercial switches~\cite{cisco:qos} do support hierarchical sharing
policies.  However, these policies are limited: First,
switches can only drop packets, and we need traffic admission control
to deal with malicious traffic.  Second, switches only offer a
link-centric sharing policy.  In practice, we desire
an \emph{end-to-end}
view for
allocating bandwidth, regardless of the physical topology.

\blue{The line of work on Dominant Resource Fairness (DRF)~\cite{ghodsi2011dominant,bhattacharya2013hierarchical}
defines a fairness metric when sharing \emph{across} multiple resource types, such as CPU, memory, and network, but rely on other mechanisms to \emph{enforce} these allocations.  \sys can be used as an enforcer for network bandwidth.}





\section{Design}
\label{sec:design}
We now detail the design of \sys, starting with
its approach to specifying bandwidth guarantees.  We then describe the
high-level system architecture, and each component in
more detail.

\subsection{Specifying Sharing Policies}
\label{sec:policies}
In \sys, a policy is specified with respect to hierarchies of services
at one contention point.  Services are traffic bundles that are
uniquely identifiable using a packet filter; a service can be a single
endpoint, a collection of endpoints, or sub-flows within an endpoint.
The service hierarchy must be a tree (i.e. no loops).
Each service has both a machine-level and rack-level policy, to
cover the primary contention points in a datacenter. A service
can have different policies for transmit and receive.

For instance,
at the machine level, all flows terminating at the Distributed File System (DFS) service port
are part of the DFS service, even if they involve multiple source
servers.  At the rack level, the DFS endpoints can be aggregated into
one service S1, and VM endpoints can be aggregated into another
service S2, but a DFS endpoint cannot be part of both S1 and S2.

A flow between a source and destination can be part of multiple
services; e.g., traffic from a MapReduce client to a DFS server
consumes bandwidth both at MapReduce and DFS, and is therefore
charged to both services.  The guarantee for traffic between two
services is limited to the minimum of the two service-specific
guarantees.  E.g., a flow between a MapReduce client guaranteed 2Gb/s
and a DFS server guaranteed 1Gb/s will be guaranteed only 1Gb/s.

A policy is configured using several parameters:
\begin{itemize}[noitemsep,leftmargin=1em,nolistsep]
  \item {\bf Min bandwidth}: the guaranteed bandwidth for the service;
    the default is 0, implying no guarantee.
  \item {\bf Max bandwidth}: the limit on the service's bandwidth;
    the default is ``unlimited.''
  \item {\bf Weight}: excess bandwidth (what remains after all
    guarantees have been met) is shared among contenders based
    on their weights (subject to their max limits).
    Weights encode a service's relative importance.
    The default weight is 1.
\end{itemize}

Policies can be specified statically, or can be changed any time to
support dynamic reservations~\cite{xie2012only}.  Note that in the
case of guarantees, admission control and job placement must ensure
that the guarantees can be satisfied in the worst case (i.e., when all
services demand their full guaranteed bandwidth).  When aggregating
services, the guarantee for the parent service must at least be the
sum of guarantees of its child services.

The most constrained policy determines the
service allocation.  For instance, consider a situation where there
are 10 MapReduce jobs in a rack, and each job has machine-level policy
(weight = 1, max = 1Gb/s).  If MapReduce jobs in aggregate have
a rack-level policy (max = 5Gb/s), and all jobs are active, each job
can send at 0.5Gb/s.  However, if only one job is active, it
cannot grab the entire 5Gb/s, since its machine-level policy
is the most constrained, and takes effect.
Thus, besides the static policy, there
is a dynamically computed \emph{runtime policy} which is actually
enforced.  Note that service policies are implicitly constrained by
machine and rack capacities.

{\bf Why have hierarchical policies?}  A hierarchical policy is one
way to quantify the over-provisioning \emph{risk}, which makes sharing
\emph{unambiguous} at every contention point, instead of letting
services battle it out between themselves.  Without hierarchy, we
would have to provision for the worst case simultaneously at each
contention point, which usually does not happen often, and can waste
capacity.  Hierarchies are also useful from an operational
perspective; since \sys exposes both guarantees and runtime policies
for every service, it is easier to debug when a service does not get
its bandwidth.

\subsection{Design components}
\begin{figure*}[t]
\techreport{\centering\includegraphics[width=0.95\textwidth]{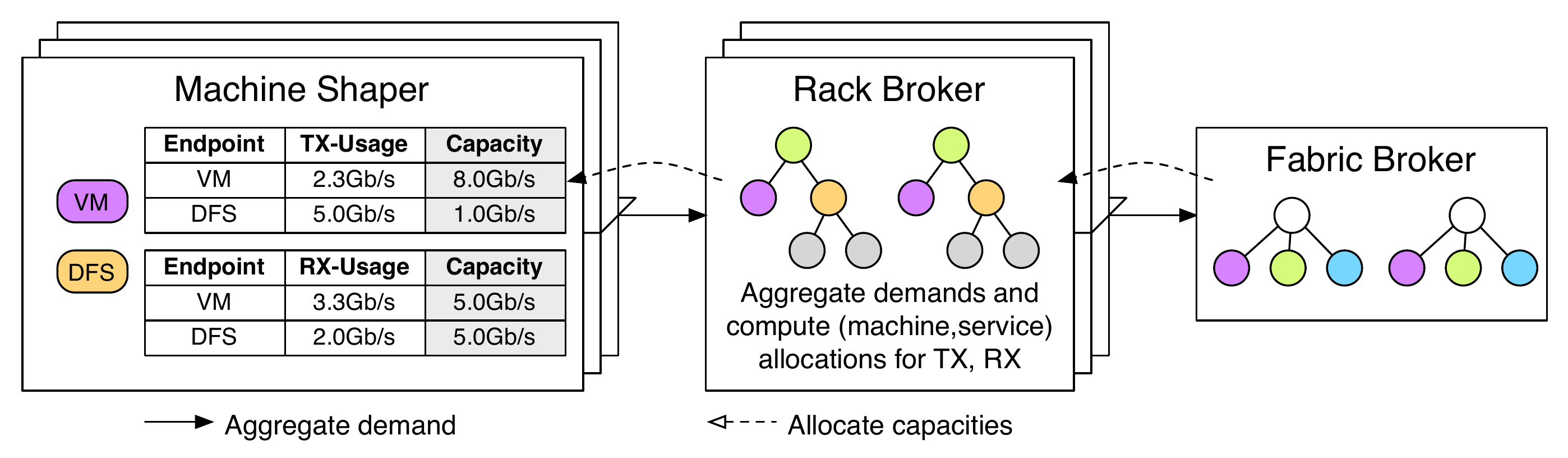}}
\nontechreport{\centering\includegraphics[width=0.8\textwidth]{figures/arch}}
  \caption{{\normalsize \sys consists of distributed machine shapers,
      rack-local brokers and a fabric broker that work together to
      enforce policies at different contention
      points.}}\label{fig:architecture}
\end{figure*}

\label{sec:components}
\sys consists of three components, each responsible for computing and
enforcing both transmit and receive bandwidth allocations, at
different contention points.  Figure~\ref{fig:architecture} shows the
high-level architecture.  These components work together to enforce
the static policy.

\begin{itemize}[noitemsep,leftmargin=1em,nolistsep]
  \item The {\bf Machine shaper} handles host fanin
    and fanout.  Its job is to enforce service shares only at a
    machine granularity, according to the static and runtime
    machine policy.  It allocates bandwidth at the
    granularity of (source,destination) pairs of communicating
    hosts~\cite{jeyakumar2013eyeq}.
  \item The {\bf Rack broker} handles the downlink and uplink overload on every rack.  Its job is to dynamically
    compute per-service per-machine policy, so that the static rack
    policy is enforced by the machine shaper.
  \item The {\bf Fabric broker} handles distributed
    rate limits, to enforce bandwidth caps on various services at the
    global scale.  It dynamically computes a per-rack,
    per-service policy to enforce the static fabric policy.
\end{itemize}

Figure~\ref{fig:example} in~\S\ref{sec:intro} shows an example with
two services: VMs and DFS.  To simplify discussion, we assume there are
no fabric-level shares for both services.  There are two machines with
10Gb/s capacities under the rackswitch, which has a 10Gb/s bandwidth to the
fabric.  The rack sharing policy is to divide bandwidth such that VMs
gets at most 1Gb/s, and DFS gets at least 6Gb/s.  Thus, when every
service is active, (M1,VM), (M2,VM) will get 0.5Gb/s each, and
(M1,DFS), (M2,DFS) are allocated 4Gb/s each.  When (M2,DFS) is idle,
(M1,DFS) is allocated 8Gb/s, and when all VMs are idle, (M1,DFS) is
allocated the entire 9Gb/s.

Note that the rack broker only computes (machine,service) shares, and
is not concerned with traffic patterns that change frequently.
The machine shaper works at fast timescales, to ensure machine runtime policy is
enforced regardless of fine-grained communication patterns.  This
timescale requirement has a direct impact on the design of the
individual components, which we detail next.

\subsubsection{Machine shaper}\label{subsec:machineshaper}
A machine may run a number of services, each perhaps in
its own virtual machine.  The problem of enforcing machine-level
bandwidth shares is the same problem solved by
EyeQ~\cite{jeyakumar2013eyeq}: On the
transmit side, the machine shaper consists of a single root rate
limiter that enforces an aggregate rate limit on the service's traffic
leaving the machine.  In addition, per-destination rate
limiters enforce receive-bandwidth shares.  The receive-bandwidth
shares are computed at the receiver, and signalled back to the sources
using a feedback packet.

On the receive side, each service is attached to a rate meter, which
is allocated some capacity $C$.  The rate meter periodically measures
the aggregate bandwidth utilization $y(t)$ of the service, and
continuously computes \emph{one} rate $R(t)$ such that
the aggregate utilization $y(t)$ matches the service capacity $C$.
This is done iteratively using a control equation:
\[
R(t+T)=R(t)\times\left(1-\alpha\frac{y(t)-C}{C}-\mathds{1_{\rm
    marked(t,t+T)}}\frac{\beta}{2}\right)
\]
\noindent where $T$ is a configurable time period, $\alpha$ is a
parameter chosen for stability.  The value $\beta$ is the fraction of
Explicit Congestion Notification (ECN) marked packets, and is used
only if there are marked packets in the interval $t$ and $t+T$
(denoted as $\mathds{1_{\rm marked(t,t+T)}}$).  The control equation
water-fills $R(t)$ such that the utilization $y(t)$ matches capacity
$C$.

The receiver samples incoming packets (1 every 10kB) and sends
feedback to the source IP address.  Sampling ensures that heavy
hitters are rate limited.  The feedback message encodes $R(t)$
serialized in a custom IP packet.  When a sender service receives
feedback from a particular destination, it creates/updates a rate
limiter to the destination, under the sender's root rate limiter, to
enforce the bandwidth allocation \techreport{(as shown in
Figure~\ref{fig:machine-shaper})}.
\techreport{
\begin{figure}[t]
  \centering\includegraphics[width=0.4\textwidth]{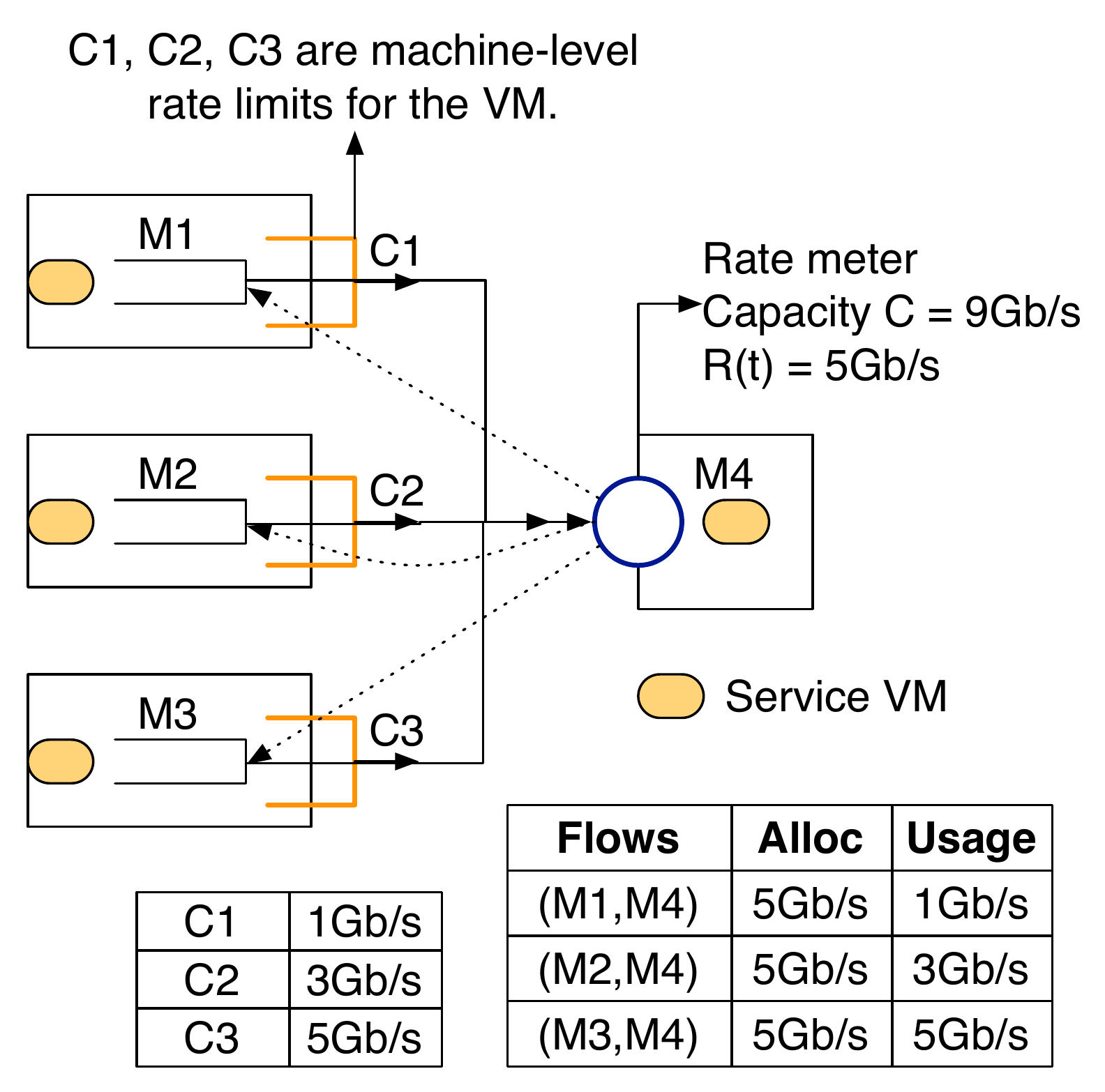}
  \caption{{An example of a max-min fair allocation (across senders at
    a receiver) computed by the machine shaper.  The allocation is
    computed at the receiver, and communicated to senders using
    feedback messages.  The end-to-end allocation is computed in a
    distributed fashion using only local
    information.}}\label{fig:machine-shaper}
\end{figure}}

{\bf Weighted allocation across senders.}
When two services communicate with each other, their receivers can divide
allocated capacity in a weighted fashion across senders: The sender
simply scales its feedback to $w_{\rm sender}\times R(t)$.  The weight
ensures that the rates of senders 1 and 2 are always in the ratio
$w_1:w_2$.

\begin{figure}[t]
  \centering\includegraphics[width=0.4\textwidth]{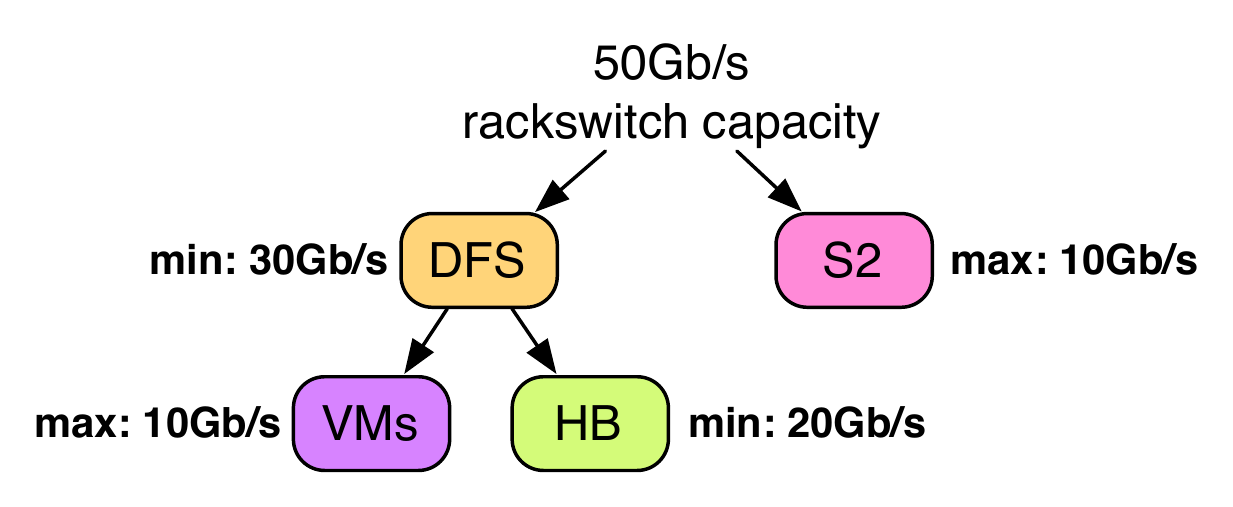}
  \caption{{Inter-tenant sharing is a special case of hierarchical
    sharing.}}\label{fig:inter-tenant}\myvspace{-1em}
\end{figure}

{\bf Inter-tenant sharing.}  For the heavy bandwidth consumers such as
DFS, a common requirement is to share bandwidth among its own users,
according to some specified policy (typically weighted shares).  Since
the notion of a `service' is flexible, it can further be
\emph{deaggregated} into `service:user.'  For instance, if HBase (HB)
and VMs both use DFS, we can create the service hierarchy shown in
Figure~\ref{fig:inter-tenant} (S2 being another service).
Deaggregation enables delegation, where services can use \sys to
manage their bandwidth.  \blue{Each packet is associated with a leaf node identified
using packet classification rules installed on the machine, both on the transmit
and the receive path.}


{\bf Parameter guidelines}: We use the same parameter settings
as in EyeQ:
$T=200\mu{}s$ and $\alpha=0.5$.  Note that the receiver \emph{does not
  keep track of the number of senders}.  This is by design: datacenter
measurements~\cite{benson2010network} show high
flow-arrival rates at a host (hundreds/second), which makes
per-flow tracking expensive and complex.

\subsubsection{Rack broker}\label{subsec:rackbroker}
The machine shaper realizes an allocation using only information
locally available at a machine, so it can only satisfy the machine
policy.  The rack broker aggregates service-level bandwidth usages
across machines in a rack, uses these values as `demands,' and
computes a machine-level transmit/receive runtime policy for each
service.  This policy determines the service's transmit and receive
capacity.  The transmit capacity is set on the (machine,service) root
rate limiter, and the receive capacity is set on the (machine,service)
root rate meter.  Once these capacities are set, the machine shaper
enforces these capacities continuously.  \emph{Note that the rack
  broker does not track (source-ip,destination-ip) communicating
  pairs.}  This is the key difference between our mechanism and a
similar control mechanism in Oktopus~\cite[\S4.3]{ballani2011towards}.

\begin{figure*}[t]
\techreport{\centering\includegraphics[width=0.95\textwidth]{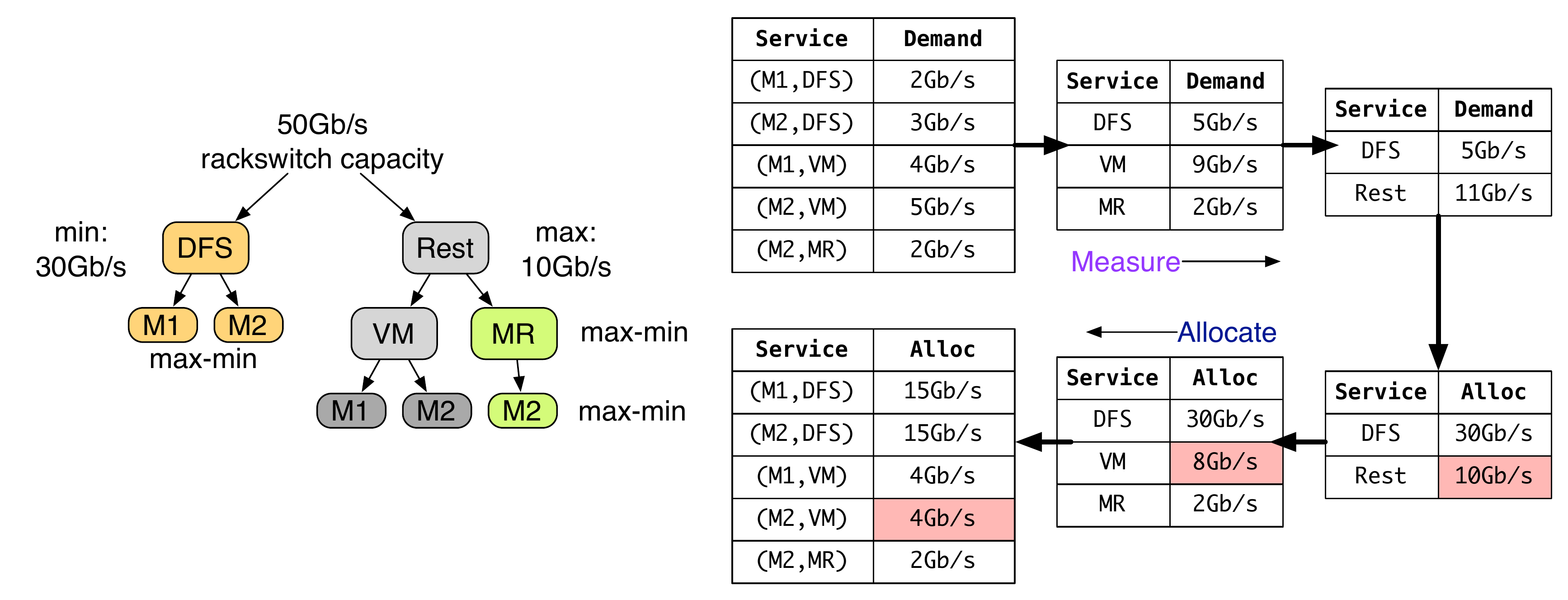}}
\nontechreport{\centering\includegraphics[width=0.8\textwidth]{figures/rack-shaper.pdf}}
  \caption{{An example illustrating the hierarchical max-min allocation,
    computed at the rack level in two steps: (i) aggregate
    measurements, and (ii) enforce allocations.  The red boxes denote
    the only entities (service instances) that are
    rate-limited.}}\label{fig:rack-broker}\myvspace{-1em}
\end{figure*}

The design of the rack broker is best explained by an example.
Consider the desired sharing policy shown in
Figure~\ref{fig:rack-broker} (assume for now that the demands were
computed before the rack broker kicks in).  The problem of determining
the per-machine service policy that satisfies the rack policy can be
done in two passes:
\begin{enumerate}
  \item A bottom-up pass from child to root that computes the
    aggregate demand at each intermediate node in the sharing hierarchy.
  \item A top-down pass that
    computes the new runtime policies at each node in the sharing
    hierarchy, according to the parent's policy.
\end{enumerate}


We make two assumptions: First, knowing the rack's capacity to the
fabric core (i.e., the 50Gb/s in Figure~\ref{fig:rack-broker}) greatly
simplifies bandwidth allocation.  This is feasible as datacenter
operators have full control over their network infrastructure, and so
the rack broker can periodically query (say) an OpenFlow controller
about the rack uplink capacities.  Second, we assume that the uplinks
are evenly utilized, so we can treat them as one single link.  \ignore{We
observe that this is indeed the case across all our datacenter rack
switches.}  But recall that the machine shaper's allocations fall back
gracefully even if there is in-network congestion (which is often only
transient).

\ignore{
\begin{figure}[h]
  \centering\includegraphics[width=0.2\textwidth]{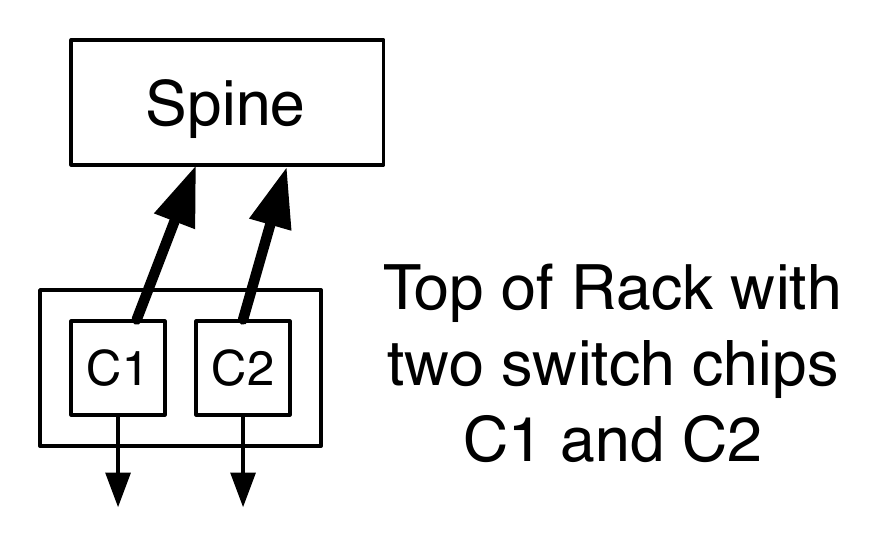}
  \caption{A high port density rackswitch with two internal chips is
    considered as two racks by the rack broker.  Note that there is no
    chip--chip communication.}\label{fig:congestion-domains}
\end{figure}

An important point is, though we term the agent a `rack' broker, there
could be multiple rack brokers per `logical' top of rackswitch. Newer
rackswitches support dense deployments of about 90--100 machines per
rack, and they are internally made of multiple smaller switches.  For
our purposes of bandwidth sharing, \emph{each smaller switch} is
considered separately for computing allocations:
(Figure~\ref{fig:congestion-domains}).
}

{\bf Allocation Algorithm}: The runtime weighted max-min allocations
and min/max guarantees can be computed by the classical water-filling
algorithm~\cite[\S~6.5.2]{bertsekas1992data}.  Each iteration of the
algorithm satiates the demand for one service.  We do not rate limit
endpoints whose demand is
less than its capacity determined by the water-fill algorithm, as rate
limiting shows a flow's progress, and increases its flow completion
time~(\S\ref{sec:latency},\S\ref{sec:lessons}).

\ignore{
{\small
\begin{lstlisting}
def max_min(demands,capacity,weights,limits):
  N = len(demands)
  W = sum(weights)
  demands = pointwise_min(demands, limits)
  total_demand = sum(demands)
  if total_demand < capacity:
    # Do not install rate limits
    return pointwise_min(capacity, limits)
  iteration = 0
  available_capacity = capacity
  final_allocations = [0, 0, ..., 0] # N times
  while iteration < N
      and available_capacity > 0:
    iteration += 1
    active_weight = sum_(i=1 to N)([weights[i] s.t. demand[i] > 0])
    share = available_capacity/active_weight
    allocations = pointwise_min(demand, [forall i: weight[i]*share])
    final_allocations += allocations
    demands -= allocations
  return final_allocations
\end{lstlisting}}
}

\subsubsection{Fabric broker}\label{subsec:fabricbroker}
The rack broker allocates bandwidth based on a capacity assigned to
the rack, and based on (rack,service) limits.  These limits can in turn
be adjusted based on the global fabric bandwidth consumption of a
particular service.  The rack broker at each rack communicates the
(rack,service) demands to a global fabric broker, which uses these
demands and in turn computes new (rack,service) allocations using the
same max-min algorithm shown above.  The fabric broker operates at a
slower timescale, running every $T_{\rm fabric}=10$ seconds.

\subsection{Scalability} The EyeQ~\cite{jeyakumar2013eyeq} paper discusses
 scalability of the machine shaper in detail.  We therefore focus on
 the rack and fabric brokers, both of which have the same design.

Using measurements from production clusters, we find that the approach
that aggregates (machine,service) tuples across machines under a rack
and fabric scales well.  To make the right allocations, the rack
broker only needs to know the bandwidth utilization of the heavy
bandwidth consuming services.  If we define a ``heavy rack-consumer''
as a service using at least 100Mb/s of rack bandwidth (averaged over
5-minute intervals) we observe 10--100 of such heavy consumers per
rack.  If we set a higher threshold of 1Gb/s for a ``heavy
fabric-consumer'' the number rapidly diminishes.  It is therefore
practical to collect and process service-level usages.



\subsection{Optimization Decomposition}
\label{sec:decomp}
The key idea behind splitting the problem into multiple sub-problems is
the principle of optimization decomposition.  A sharing objective is
decomposed into multiple sub-problems that execute in a distributed
fashion.  We refer the reader to~\cite{palomar2006tutorial} for a
detailed discussion on the mathematical foundations of optimization
decomposition.

\begin{figure}[h]
  \centering\includegraphics[width=0.4\textwidth]{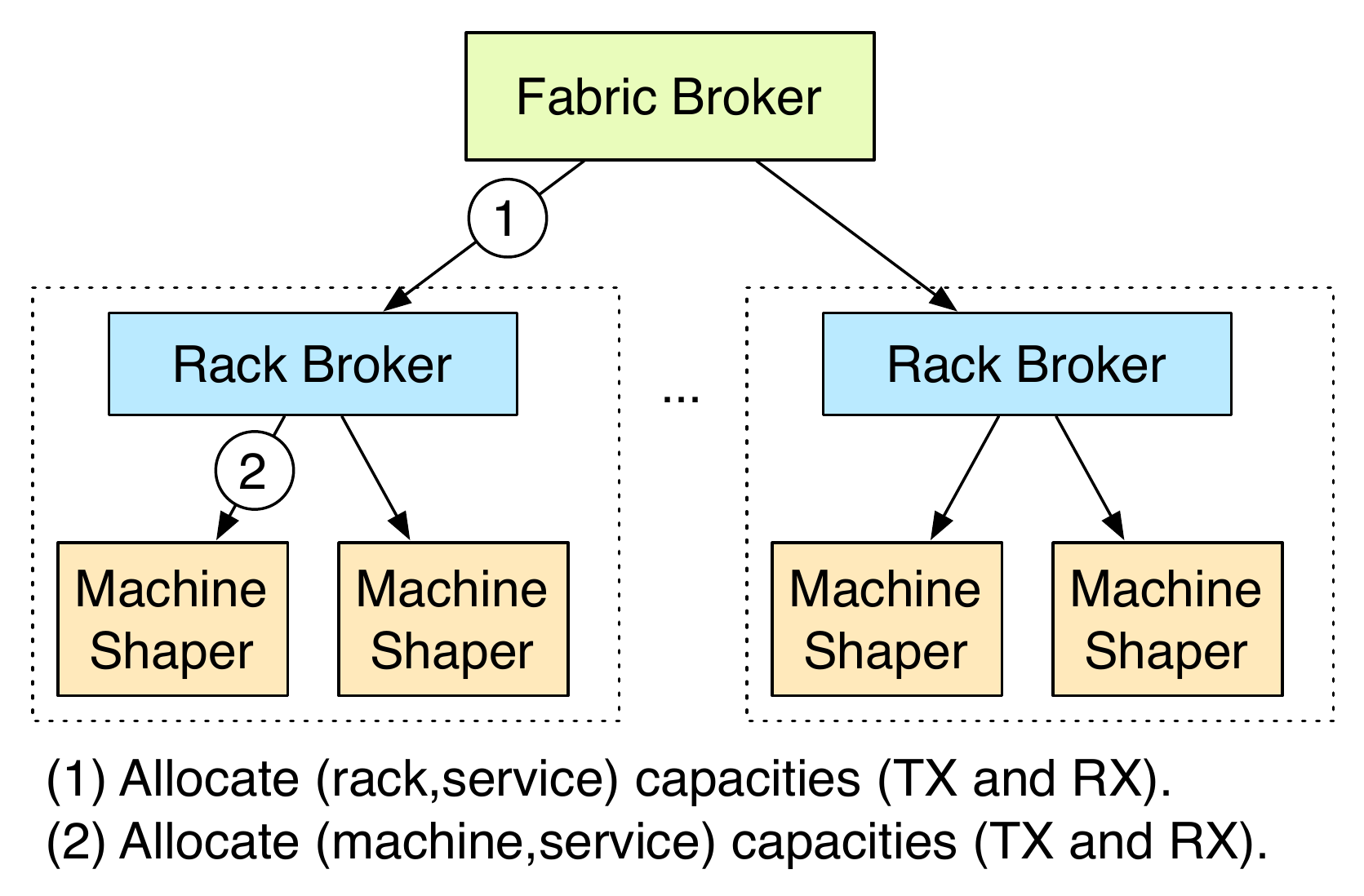}
  \caption{Optimization decomposition in \sys.}\label{fig:opt-decomp}
\end{figure}

\sys uses this approach, which
has also been adopted by other
bandwidth-management
systems~\cite{jain2013b4,shue2012performance}, and inherits several
benefits:

\begin{itemize}
  \item {\bf Scalability}: Each shaper/broker is responsible only for
    a portion of the datacenter, and the optimization decomposition
    leads to a design that scales with the datacenter size.  The
    machine shaper deals with no more than a few tens of services;
    the rack broker deals with no more than a few
    100s of (machine,service) pairs in the rack; and the fabric
    broker deals with more no than tens to hundreds of
    racks.
  \item {\bf Fault tolerance}: The scale-out design ensures graceful
    degradation, as un-failed parts of the system will continue normal
    operation.  For instance, since the machine shaper uses
    ECN marks to detect when
    the network is congested, the rack broker can
    fail without significantly affecting the system.  The allocations degrade
    gracefully, and by design, the receivers will get a max-min share
    of the rackswitch bandwidth.
\end{itemize}

The main disadvantage of optimization decomposition is that the
bandwidth allocation according to specified policy is not computed in
one shot, but over a sequence of iterations.  This can delay the speed
at which a service can get its full allocation.  While this is
typically not a big concern at fabric scale, a service must be able to
grab its fair bandwidth share quickly at the rack level.  Our design
choice to not rate limit services that are not consuming their full
bandwidth share gives them the ability to ramp up quickly.

One possible concern is that services, when not rate limited, can
burst to gain an unfair share of bandwidth and affect other service
allocations.  The effect of this burst is limited, as the
machine-shaper kicks in quickly (within hundreds of micro-seconds)
during transient network congestion, ensuring that services are not
starved.

\subsection{Are demands stable?}
\begin{figure}[h]\myvspace{-1em}
  \centering\includegraphics[width=0.45\textwidth]{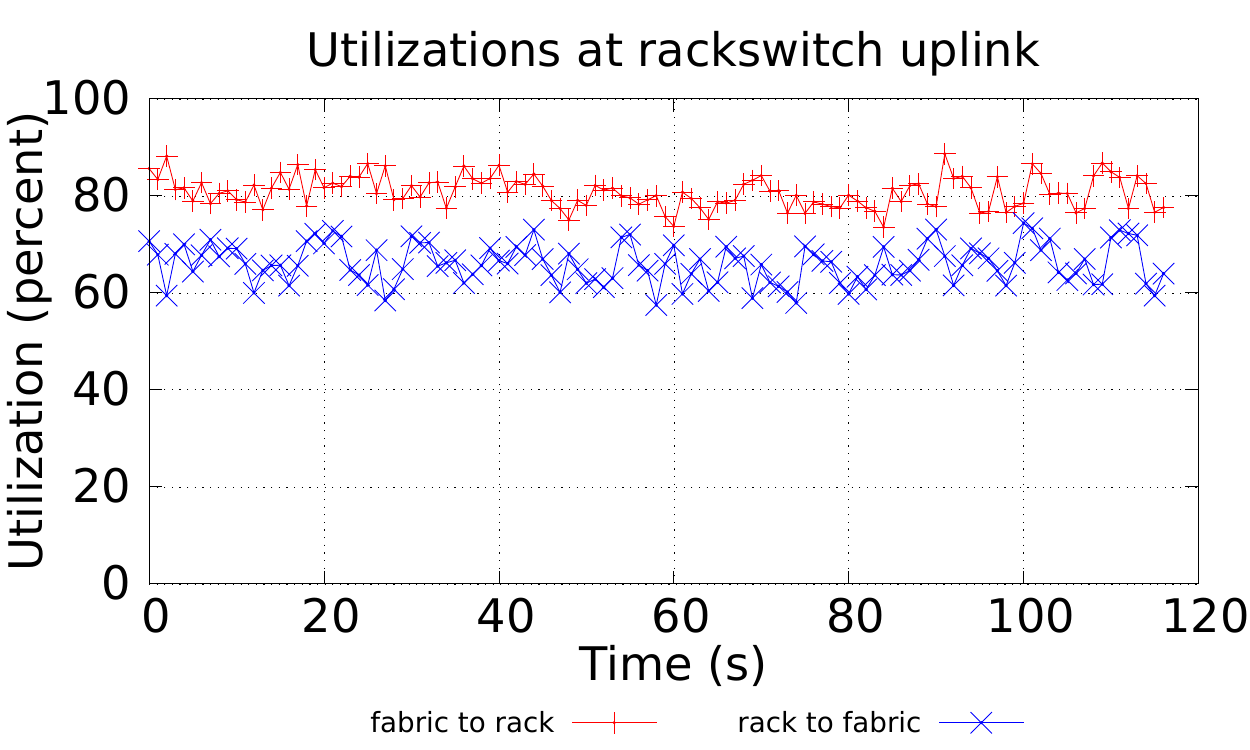}
  \caption{{Time series of uplink and downlink utilizations (averaged
    over 1s) of a highly utilized rack, showing that the link
    utilizations are fairly stable over multiple
    seconds.}}\label{fig:high-util-rack}\myvspace{-1em}
\end{figure}

The bandwidth brokers operate at three different timescales: (i)~the
machine shaper operates at network round-trip timescales, (ii)~the rack broker
operates at timescales of seconds, and (iii)~the fabric broker
operates on multiple tens of seconds.  One might wonder if the
(machine,service) demands are stable for a long enough period
that the rack broker can kick in, and the same for the global fabric
broker.  Figure~\ref{fig:high-util-rack} presents evidence that demands are
indeed stable, so the individual brokers can operate effectively over
the timescales they are designed for.

By design, the rack broker kicks in and enforces allocations only when
the bandwidth usage is above limits specified by policy.  \ignore{In
  most racks, the top bandwidth consumer is the distributed file
  system traffic, consuming as much as 60\% of the rack bandwidth.}
Figure~\ref{fig:high-util-rack} shows the utilization (averaged over
1 sec.) of one highly-utilized rack \blue{in a production data center}.
We see that the aggregate
transmit and receive utilizations remain fairly stable over multiple
seconds, and thus the timescale of a few seconds is pragmatic.  \blue{We don't
have visibility at smaller timescales in production
traffic.}\ignore{Though
the utilization does not always touch 100\%, an adversary, or buggy
application can easily saturate the uplinks.
\fixthis{I don't understand the motive for the prior sentence.}
}

\section{Provisioning for Latency}\label{sec:latency}
In \S\ref{sec:background}, we showed how a bandwidth guarantee enables
services to achieve a bound on the tail flow completion times.  The
formula, however, made an assumption that flow arrivals are Poisson,
and that flow sizes are exponentially distributed with a given mean.

While this model serves as a useful guide to provision systems, it is
often the case that flow arrivals are not Poisson.  Prior datacenter
measurement work shows that flow arrivals are
\emph{bursty}~\cite{benson2010network}.  Therefore, we seek a
guideline that does not make assumptions on flow sizes (except that
they are finite), or flow service order (except that the system is
work-conserving), or any particular flow arrival pattern.  Since we
make minimal assumptions, the bound will serve as a useful guideline
to reason about the worst-case performance.

We model a latency-sensitive endpoint as a work-conserving queue of
capacity $C$ that is shared with other endpoints.  It is impossible to
have any guarantees unless we constrain the arrivals in some fashion.
Suppose $B(t_1,t_2)$ refers to the number of bytes received by the
queue between time $t_1$ and $t_2$.  Consider a constrained arrival
process parameterized by
$(\sigma,\rho)$~\cite{cruz1991calculus} such that:
\begin{equation}
\label{eqn:sigmarho}
B(t_1,t_2) \leq \sigma + \rho\;C\;(t_2-t_1)
\end{equation}

\noindent for all $t_1<t_2$.  Here, $\sigma$ quantifies the arrival
burstiness, and \mbox{$\rho\in(0,1)$} quantifies the average long-term
rate.  Then, we can show that the flow completion time ${\rm FCT}(f)$
of any flow $f$ of size $Z(f)$ satisfies:

\begin{equation}
\label{eqn:result}
{\rm FCT}(f) \leq \frac{\sigma+Z(f)}{C(1-\rho)}
\end{equation}

\techreport{
The proof is by a counting argument: Let $S(f)$ and $F(f)$
denote the start time and finish time for a flow $f$.  Now, the number of
bytes arriving to the server between $S(f)$ and $F(f)$ is constrained
by Equation~\ref{eqn:sigmarho}.  We have:

\begin{proof}[\nopunct]\renewcommand{\qedsymbol}{}
\begin{align*}
F(f) & \leq \text{Start time + Waiting time + Service time} \\
F(f) & \leq S(f) + \frac{B(S(f),F(f))}{C} + \frac{Z(f)}{C}\\
(1-\rho)\;F(f) & \leq (1-\rho)\;S(f) + \frac{\sigma+Z(f)}{C}
\end{align*}
Since $\rho\in(0,1)$, the quantity $(1-\rho)>0$, and therefore:
\begin{align*}
F(f) - S(f) & \leq \frac{\sigma+Z(f)}{C(1-\rho)}\\
\implies \;\; {\rm FCT}(f) & \leq \frac{\sigma+Z(f)}{C(1-\rho)}
\end{align*}
\end{proof}}

\iftechreport\else
The proof is by a counting argument and we present it in our technical
report~\cite{techreport} due to space constraints.
\fi

\emph{Alternatively, this bound can be used as a guideline to limit
  the peak load $\rho$ on the network to keep FCT under a desired
  value.}  The bound only assumes a $(\sigma,\rho)$ constraint on
arrivals to the work-conserving queue, and that latency-sensitive
flows are not rate limited.  A rate-based congestion control can
control the long-term load $\rho$.  The burst size $\sigma$ is the
maximum of two components: (i) the excess line-rate burst due to
delayed convergence of the congestion control algorithm, and (ii) the
burstiness of sender rate limiters.  From experiments, we found that
the burst due to convergence time is the dominant factor; i.e., if the
congestion control algorithm converges in $t$~seconds, then the queue
sees a maximum burst of $C\times{}t$.  \ignore{We elaborate on this in
  our technical report.}

\techreport{

For example, if C=100Mb/s, and the congestion control algorithm takes
10ms to converge, then the maximum burst size is about 83 MTU sized
packets.  Once the congestion control algorithm converges, the only
burstiness is due to rate limiters at senders.  To quantify this
burst, we set up an experiment using the Mininet network emulator as
follows: 100 hosts talk to one receiver using long-lived TCP flows
rate limited to to 1/100th of the desired load $\rho$ at the receiver.
The rate limiters were configured with a maximum burst size of 64kB,
and we measure the queue sizes at the receiver.
Figure~\ref{fig:qsize} plots the distribution of queue sizes for
various values of $\rho$.  We find that even at 90\% load, the 99th
percentile queue size is less than 25 packets, which is smaller than
the maximum burst due to delayed convergence (83 packets), indicating
that the burst contribution due to the delayed convergence of
congestion control algorithm plays a dominant role in determining the
worst case flow completion time.

\begin{figure}[t]
\centering
\includegraphics[width=0.45\textwidth]{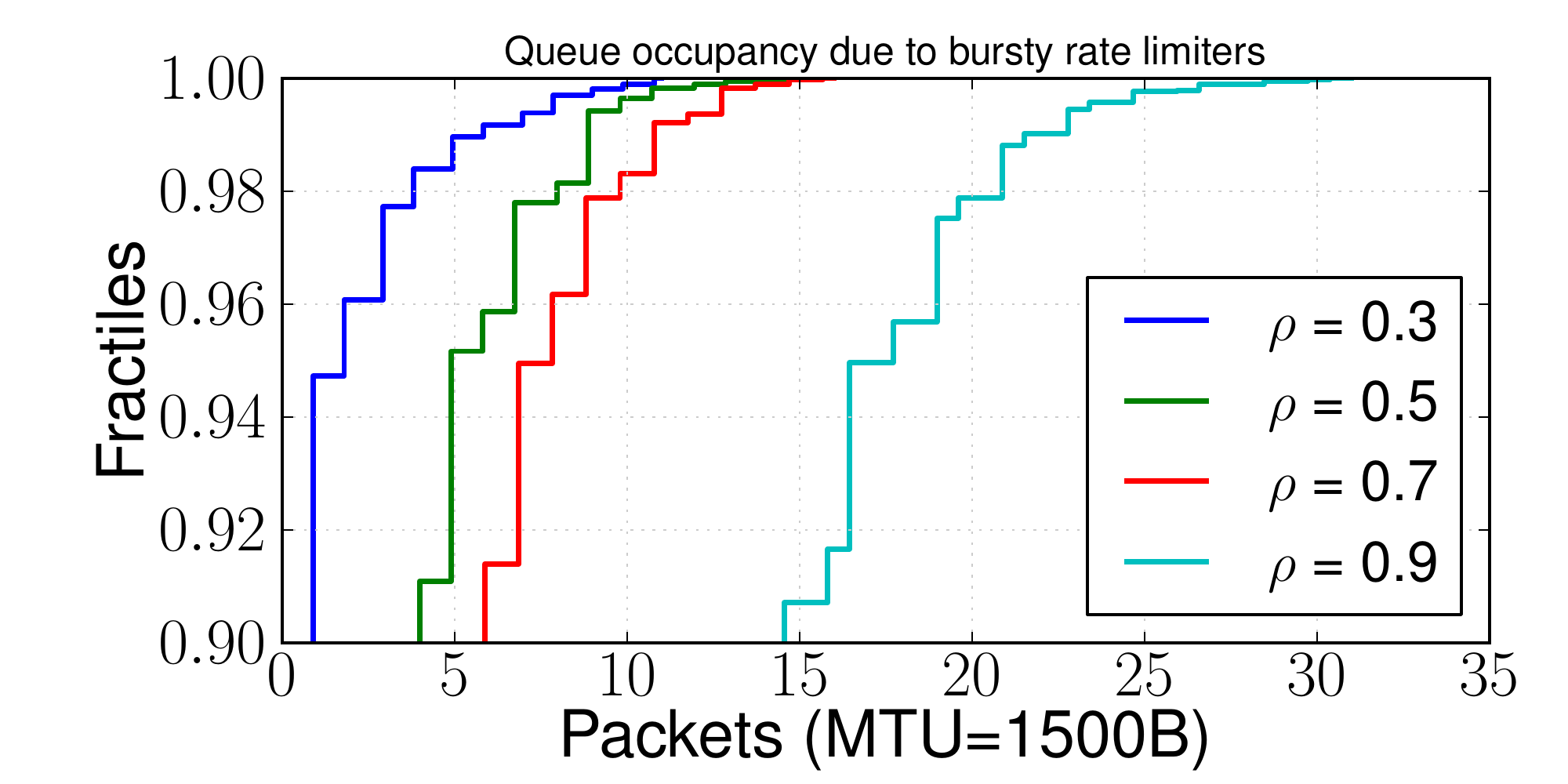}
\caption{{CDF of queue sizes as a function of load.}}\label{fig:qsize}
\end{figure}

Though the bound is on the worst case FCT, flows can be reordered
(e.g. through prioritization) to achieve better individual FCTs.  How
a service schedules its own flows in accordance to bandwidth
guarantees is completely up to it.  We echo the insight of recent work
pFabric~\cite{alizadeh2013pfabric} that rate control is perhaps a poor
choice for fine-grained prioritization across \emph{mice flows}
(e.g. flows less than a few 100kB).  \sys's rate control is decoupled
from the flow's priority.  In practice, datacenter
measurements~\cite{alizadeh2010data,greenberg2009vl2} show that most
of the bandwidth consumption is from the elephant flows that last long
enough to be accurately rate limited (flows between a few MB to 100s
of MB last a few msec. to 100s of msec.).  }

\section{Implementation}
\label{sec:implementation}
Our prototype of \sys builds on top of EyeQ.\footnote{\url{http://jvimal.github.io/eyeq/}}
We first briefly review the implementation of
the machine shaper in EyeQ, discuss our new improvements, and then
describe the implementation of the rack broker and fabric broker.
Though we prototyped \sys in software, its
control logic (rate brokers) could interface with a hardware datapath
(e.g., rate limiters on a NIC, or in switches within the fabric).

\subsection{Machine shaper}
\begin{figure}[h]
  \centering\includegraphics[width=0.35\textwidth]{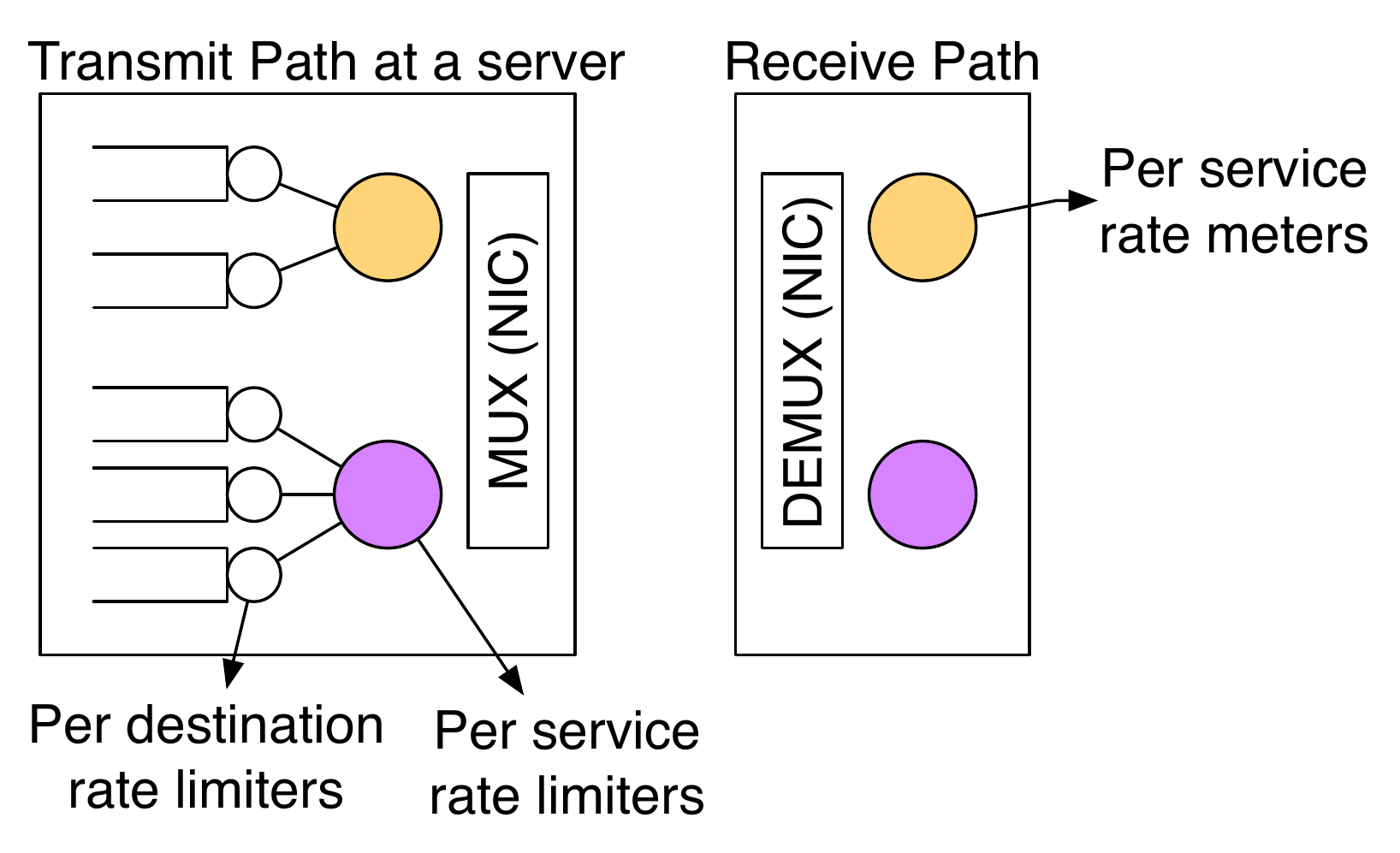}
  \caption{{The rate limiter hierarchy at the sender, and rate meters
    at the receiver, shaded by service type.}}\label{fig:rl-hierarchy}
\end{figure}

The machine shaper consists of a transmit and receive side component.
As in EyeQ, we use hierarchical rate limits to enforce sharing
policies on the transmit side.  At a service endpoint on a machine,
the rate limiter hierarchy is shown in Figure~\ref{fig:rl-hierarchy}.
In the hierarchy, rack uplink traffic is controlled by rate limits set
by the rack broker (on the service root rate limiter).  This means
that inter- and intra-rack traffic share the same rate limit if there
is congestion at the rack uplinks; however, when we look at
measurements from several datacenters, we see that most of the traffic
is \emph{inter-rack}, so we did not optimize for fate sharing.
Nevertheless, we could separate inter- and intra-rack traffic below
the service root rate limiter, and have the rack broker set limits
only on inter-rack traffic.  The rate limiters and rate meters are
designed for high throughput using per-CPU datastructures.

\myvspace{-1em}

\subsection{Rack broker}
We implemented the rack broker as a user-space program in C++.  There
are many ways to query the aggregate link utilization across services:
the program can install counters on the rackswitch and use
OpenFlow~\cite{mckeown2008openflow} to count the per-service
utilization.  However, for ease of deployment, we implemented a 
simple approach to aggregating the (machine,service) counters.  The
rack broker is distributed across machines, and each machine queries
the local machine shaper to obtain a list of (machine,service) pairs.
Each machine then broadcasts this list of tuples to other machines in
the rack, and aggregates tuples from other machines.  Thus, each
machine will have its own copy of the list of (machine,service)
counters for all machines and services in the rack.  Using this
information, each machine runs the water-fill algorithm to determine
all (machine,service) allocations, and each machine locally enforces
allocations for its own services.  The machines broadcast the counters
using UDP once every $T_{\rm rack}$ seconds.

{\bf Tolerance to failures}: In the absence of rack shaper, the
machine shapers fall back gracefully to a max-min fair allocation
across receiving VMs/service instances (as in EyeQ).  Thus, rack
broker failures do not cause global failures.  However, the rack
broker is highly fault tolerant, as it runs on every machine.  If a few
machines die, the remaining rack brokers continue working as expected.
If the broadcast packets are intermittently lost, the last updated
value of (machine,service) is used to compute bandwidth shares.  We
prevent small rate allocations from sticking permanently by having
machine shapers reset their allocations to their static configuration
after a long timeout.  This timeout value $T^t_{\rm rack}$ is chosen
so that the machine shaper can be certain about a rack broker failure.

\subsection{Fabric broker}
The fabric broker uses the (rack,service) shares for a given rack, to
determine new (rack,service) allocations on a per-rack basis.  The
fabric broker runs relatively infrequently on a few machines spread
across the datacenter.  On each rack, one of the rack brokers is
designated as the leader, and the leader sends the rack-level
bandwidth consumption to the fabric broker's service address.  As with
rack broker, the fabric broker need not be perfectly reliable.

The rack leader sends an RPC to the fabric broker once every 10
seconds.  Even with 10000 racks within a cluster, sending 10kB data
(which encodes utilizations for over 1000 services) to the fabric
shaper consumes only 80Mb/s of traffic at the fabric broker.

{\bf Tolerance to failures}: The strategy that rack brokers use to
handle handle fabric broker failures is exactly the same as the
strategy machine shapers use to handle rack broker failures.  However,
we set a larger timeout, $T^t_{\rm fabric}=50\text{~seconds}$.
If the rack broker doesn't hear from the fabric broker in 50~seconds,
it resets the runtime policy to the statically configured policy.

\section{Evaluation}
\label{sec:evaluation}
We demonstrate that \sys achieves our goals of predictable bandwidth
sharing.

Since we built \sys on top of EyeQ, we refer the reader to
measurements in the EyeQ paper~\cite[\S5.1]{jeyakumar2013eyeq}.
\sys inherits EyeQ's results for: (i)~low CPU overhead when
compared to state-of-the-art software rate limiters;
(ii)  millisecond-timescale convergence times at the machine shapers.

The new aspects of \sys that we cover are:
\begin{itemize}[noitemsep,leftmargin=1em,nolistsep]
  \item Rack brokers are highly efficient, and consumes less than 2ms
    wall-clock time to allocate bandwidth to as many as 100k services.
    The bandwidth overhead is less than 3Mb/s when rack broker
    interval is 1s.  Moreover, they converge to the fair rates within
    a few seconds.
  \item In an emulation, the fabric broker converges within 10s to
    limit the maximum bandwidth consumption across 100 racks.
  \item The rack broker can control the maximum utilization on the
    rackswitch, ensuring predictable tail latency for services in
    the rack, even in the presence of a malicious service that tries to
    grab all bandwidth.
  \item \sys, unlike EyeQ, controls tail latency even if the network core
    is congested, while allowing work-conserving allocations to services.
\end{itemize}

{\bf Topology and parameters}: Unless otherwise noted, we use a
leaf-spine topology, where 9 rackswitches are connected to 2 spine
switches in a bipartite graph\techreport{ as shown
  in Figure~\ref{fig:topology}}.  There are 10 hosts under each of the 9
rackswitches.  Each host has a 10Gb/s NIC.  The network is
oversubscribed 1:1.25 at the rackswitch.  We use the \sys parameters
shown in Table~\ref{tab:parameters}.

\begin{table}[t]
\begin{center}
\begin{tabular}{ |p{0.25\textwidth}|r|r| }\hline
RCP aggressiveness & $\alpha$ & 0.5 \\\hline
RCP time interval & $T$ & 200$\mu$s \\\hline
Rack broker frequency & $T_{\rm rack}$ & 1s \\\hline
Fabric broker frequency & $T_{\rm fabric}$ & 10s \\\hline
Rack broker timeout at machine broker & $T^t_{\rm rack}$ & 5s \\\hline
Fabric broker timeout at rack broker & $T^t_{\rm fabric}$ & 50s \\\hline
ECN marking threshold & & 80kB\\\hline
\end{tabular}
\end{center}
\caption{Parameters used in evaluating \sys.}\label{tab:parameters}\myvspace{-4pt}
\end{table}
\techreport{
\begin{figure}[t]
\centering
\includegraphics[width=0.4\textwidth]{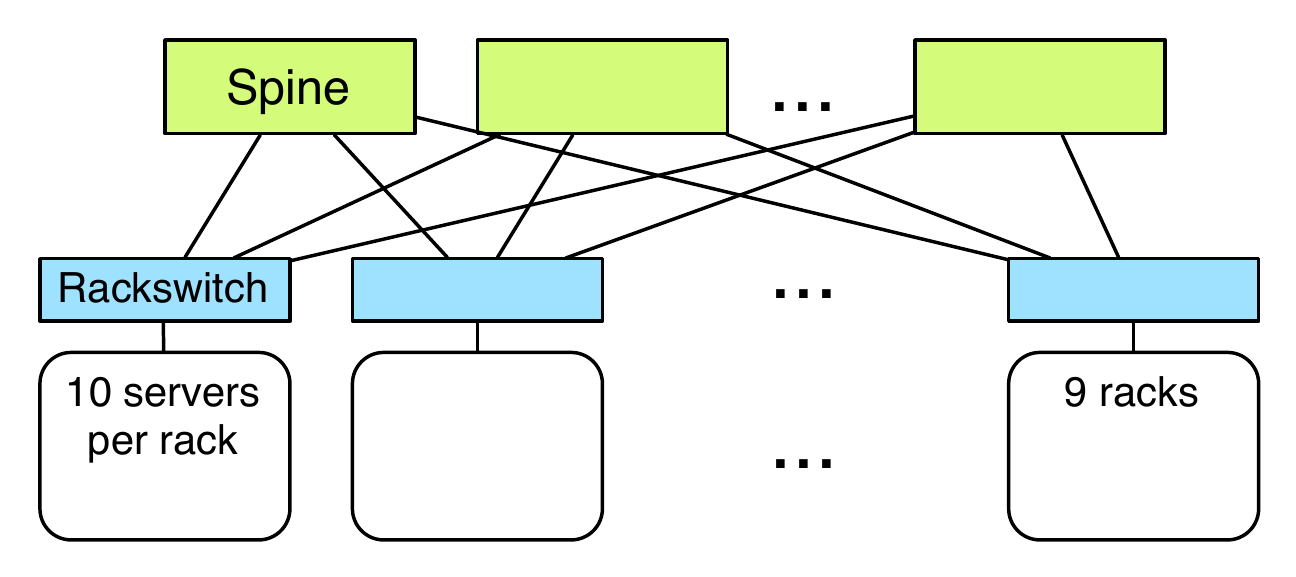}
\caption{{Topology of the testbed.  The network is oversubscribed by a
  factor of 1.25 (100Gb/s to 80Gb/s) at the
  rackswitch.}}\label{fig:topology}
\end{figure}}


\techreport{
\subsection{Machine shaper microbenchmarks}
{\bf Graceful degradation during network congestion}: The machine
shaper senses congestion when link utilization approaches capacity,
which is not sufficient if the network is congested.  Hence, the
machine shaper also uses ECN feedback from the network to back off
when links are congested.  We evaluate this scenario by having two
service instances under different racks send traffic to two their
respective receivers under the same rack, but on two different machines.
We induce network congestion by disabling all but one fabric link at the
receiving rackswitch, creating a 2-to-1 over-subscription at the
rackswitch.

Figure~\ref{fig:network-congestion} highlights an important point that
the machine shaper must operate at a fast enough timescale.  At time
t=0, only one service is active.  At t=40s, both services are active,
causing network congestion; therefore, bandwidth is shared almost
equally (Jain's Fairness Index is 0.99).  At time t=60s, we increased
$T$ from 200$\mu$s to 1ms; notice that the machine shaper is now too
slow to guarantee fair shares across receivers, as the transport layer
(TCP) also backs off due to network congestion.

If we were allowed to
change TCP to back off less aggressively, or to use UDP, a 1ms timescale
for reaction time would be sufficient.  We discuss the tradeoffs
between accurate and coarse-grained rate limiting and its impact on
flow completion times in~\S\ref{sec:lessons}.

\begin{figure}[t]
\centering
\includegraphics[width=0.45\textwidth]{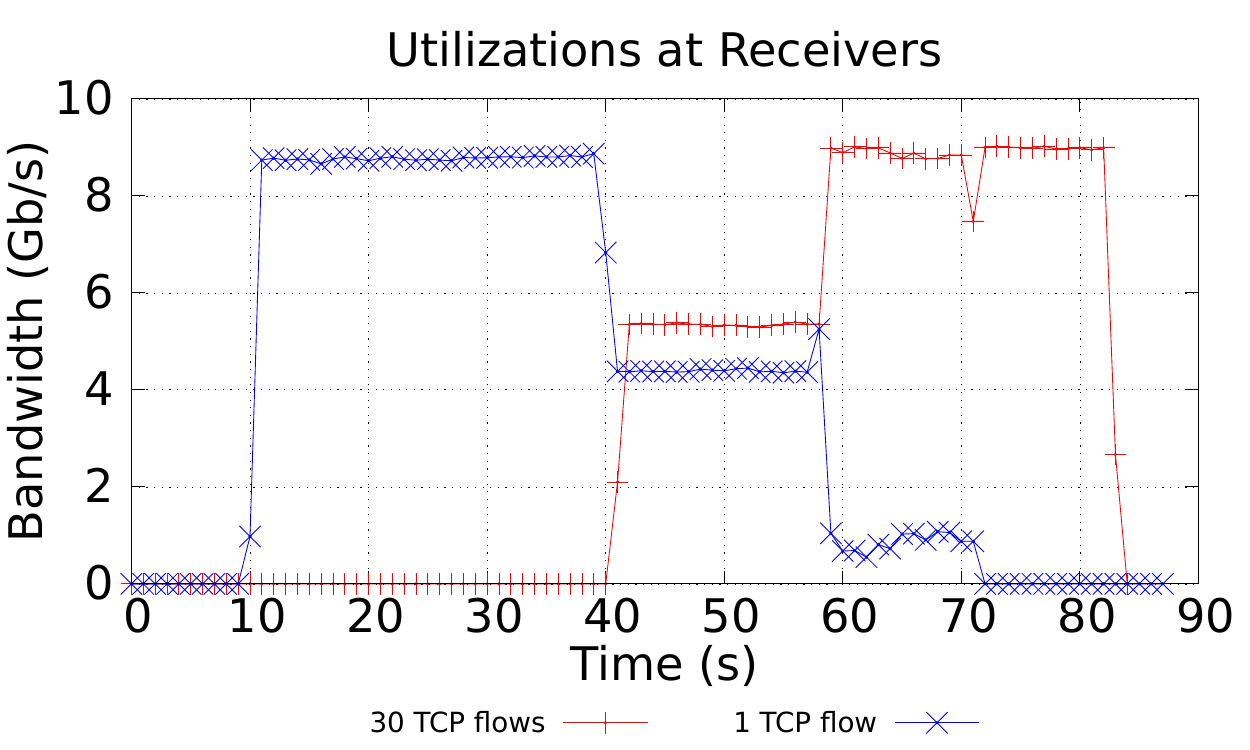}
\caption{{When there is network congestion at the rackswitch, \sys
  gracefully falls back to a sharing model where receivers get
  (almost) equal shares of the network bottleneck link.
  \ignore{However, a
  fast reaction time ensures nearly ideal sharing.
  \fixthis{I don't think the figure actually demonstrates the
  last sentence in the caption.  Remove it?}}}}\label{fig:network-congestion}
\end{figure}
}

\subsection{Microbenchmarks: Rack and Fabric brokers}
{\bf Bandwidth overhead}: In our rack broker implementation,
each machine broadcasts its own service utilizations to other machines
under the rack.  Each service can be represented as a 4B integer, and
its utilization as another 4B floating point number.  At 8B per
service, even 1000 bandwidth-intensive services in a rack would
generate only 8KB of data, plus some overhead for packet headers.
If there are 40
machines in a rack, unicasting this data to each machine once/sec
would cost less than 3Mb/s per machine.

{\bf Computation overhead}: The rack broker repeatedly computes
hierarchical max-min shares among $N$ services.
Table~\ref{tab:rackshaper-compute} shows the wall-clock time
of computing these
shares for a single level in a hierarchy, for various values of $N$
(using one core of a 2.4GHz Linux desktop machine).  We initialized
demands randomly such that the load is close to the capacity.  \blue{Although
our implementation of the water-fill algorithm~\cite[Sec.~6.5.2]{bertsekas1992data} is $O(N^2)$,
the wall-clock time to convergence scales well
with $N$.  This is because every iteration of the water-fill
satisfies the demand for \emph{at least} one service, and we track demands
with a precision of 1Mb/s.  Thus, it does not take more than $\sim$80000
iterations to water-fill 80Gb/s capacity.}

\begin{table}[h]\centering
\begin{tabular}{l|r|r|r|r}
$N$   & 100     & 1k       & 10k       & 100k \\\hline
Time  & 2$\mu$s & 12$\mu$s & 320$\mu$s & 1.6ms
\end{tabular}
\caption{Average wall-clock time per iteration of max-min share computation.}
\label{tab:rackshaper-compute}
\end{table}

{\bf Convergence speed}: The hierarchical max-min algorithm converges
in \emph{one invocation}, and therefore the convergence speed only
depends on the speed of exchanging information and the computation
time.  The rack broker can also work at the rate of 10 times per
second, consuming 30Mb/s bandwidth, converging to the allocations in
100ms.  In practice, however, we found 1s time granularity to be
sufficient in ensuring isolation, as the machine shaper works at
faster timescales to ensure the network does not experience persistent
congestion.

\begin{figure}[t]
\centering
\includegraphics[width=0.48\textwidth]{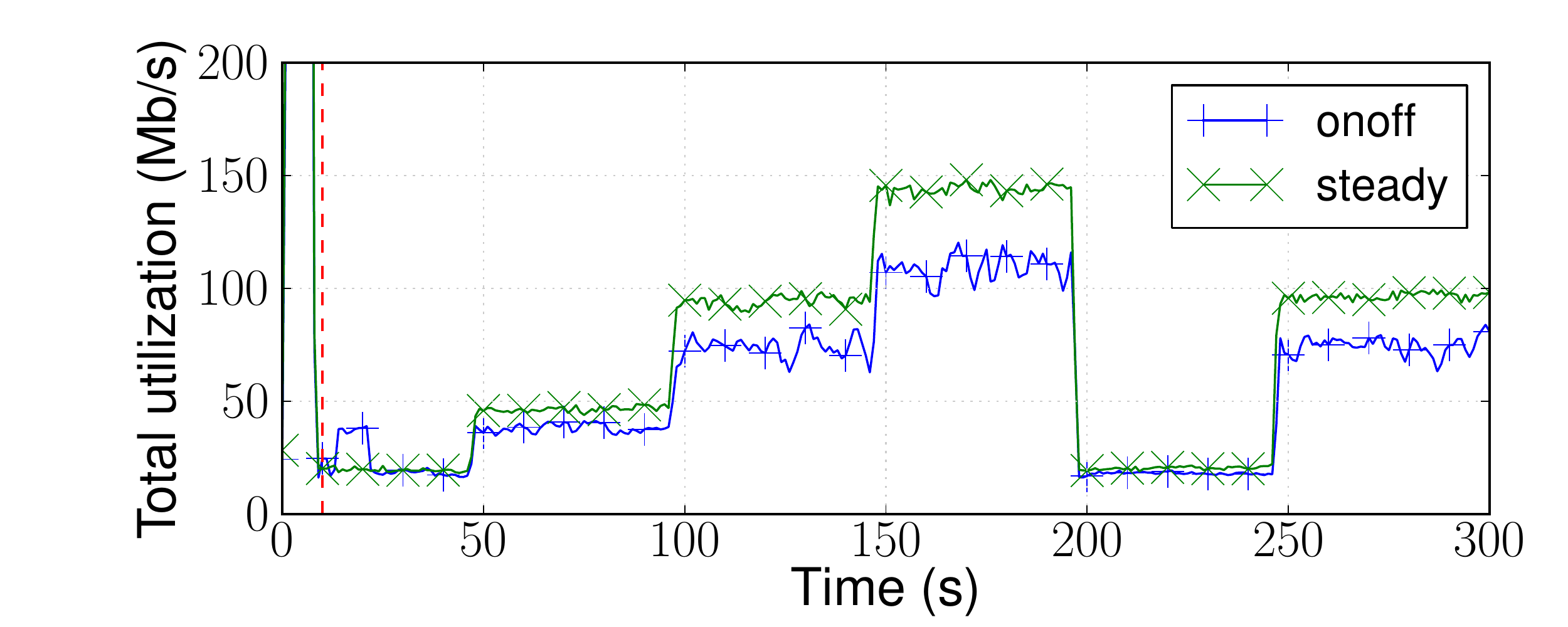}
\caption{{At t=0, we start the fabric broker, and it takes 10s
    for it to take effect. However, once the fabric broker starts to
    allocate bandwidth, it converges within a few iterations (at
    t=30s) and limits the tenant's usage to $<$20Mb/s, despite its
    bursty traffic pattern.}}\label{fig:fabric-broker}\myvspace{-1em}
\end{figure}

{\bf Fabric broker}: To test the convergence of the fabric broker at
large scale, we emulated a 100 rack cluster in
Mininet~\cite{handigol2012reproducible} with one tenant that is rate
limited to 20Mb/s.  Each rack has a 100Mb/s link to the fabric; it
picks another rack at random and sends a UDP data burst for 5s, and
sleeps for another 2s until t=300s (on-off).  In another experiment,
the traffic is less bursty (steady): it does not sleep for 2s, but
instead, it immediately sends traffic to another random rack.  The
burst time is smaller than the 10s interval at which the racks
communicate with the fabric broker.  Figure~\ref{fig:fabric-broker}
shows the time series of the aggregate utilization of the tenant: as
we can see, the fabric broker converges after the first burst, and
within a few iterations, the tenant is limited to its 20Mb/s, 50Mb/s, 100Mb/s,
    150Mb/s, 20Mb/s and back to 100Mb/s every 50s (to illustrate
    convergence).  In
practice, the rack broker will ensure the initial burst does not
adversely affect the performance of other services within the rack.
\highlight{This global service limit would not be feasible with proposals such as
EyeQ/ElasticSwitch, as the service's per-rack limit depends on
bandwidth usages of other racks.}




\subsection{Macrobenchmarks}\label{subsec:macro}
{\bf Policy}: We demonstrate how \sys can
protect the rackswitch links from persistent congestion on a real
(i.e., non-emulated) network.  On each
machine in the topology, we
instantiate two service endpoints A and B, each of which is given equal
bandwidth at the machine level.  At the rack uplink, service A is
given at most 30Gb/s,
and service B is
given at least 30Gb/s, but the peak load on the rack is limited to 60Gb/s.  Recall that \sys's final
allocation depends on the most constrained resource.

\begin{figure}[t]
\centering
\includegraphics[width=0.45\textwidth]{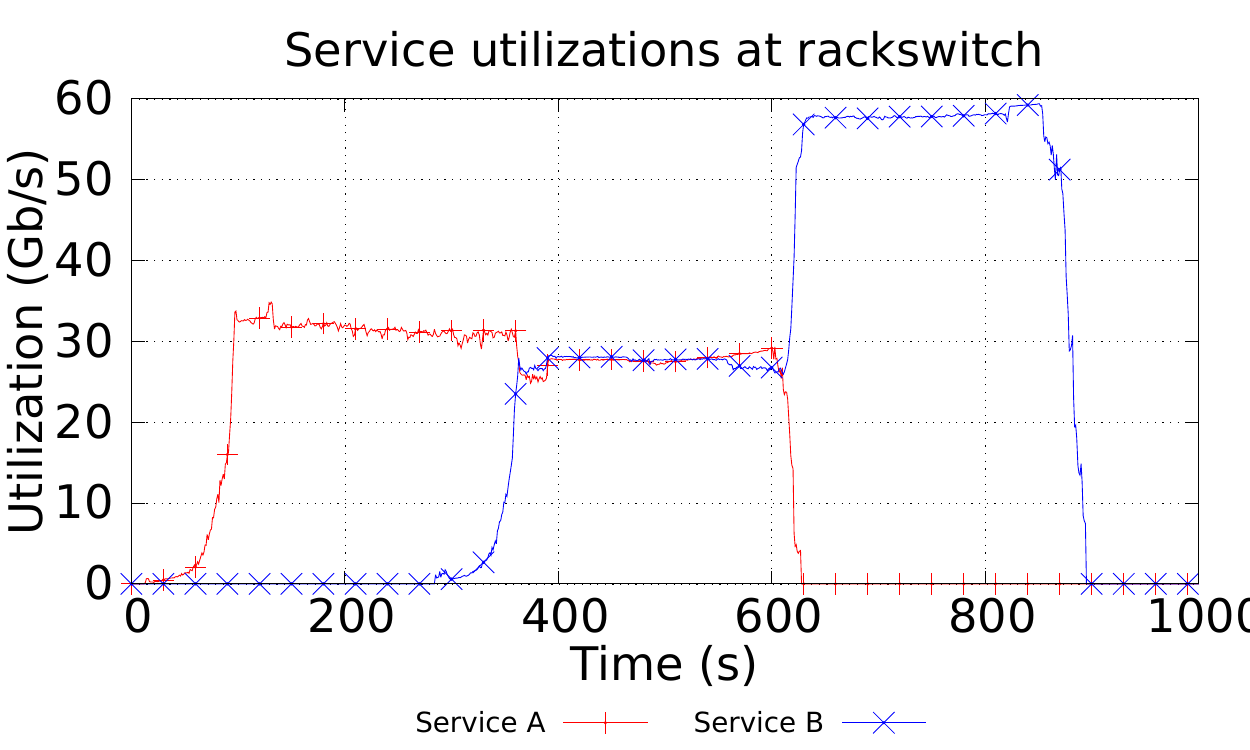}
\caption{{Two services A and B sharing limited rackswitch capacity.
  Service A can consume at most 30Gb/s, while service B can consume at
  least 30Gb/s but the total is limited to 60Gb/s.  The utilization data is obtained
  every second, and the plot marks are every 30s.
  }}\label{fig:rackswitch}\myvspace{-1em}
\end{figure}

{\bf Throughput protection}: One of the racks is designated as the
receiver.  The traffic pattern consists of long-lived transfers
between every pair of sender and receiver (a full mesh).
Figure~\ref{fig:rackswitch} shows the utilization of each
service at the receiving rackswitch.  At the start of the
experiment, only service A is active; notice that it uses only
30Gb/s.  At about T=300s,
we start service B, which
takes around 50s to ramp up.  \sys converges faster than the time it
takes for all flows of service B to ramp up, but by the time all flows
are running at full line rate, service B is able to utilize its full
30Gb/s link capacity.  When service A finishes, service B is able to
utilize all the remaining capacity up to 60Gb/s.

{\bf Latency protection}: Using the same policy as above, we show how
\sys can protect the latency of a service against high-rate adversarial
traffic.
We run the same
configuration as above, and both services mimic a client-server
workload with all-to-all communication between senders spread
across all but one rack, and receivers in the one remaining rack.
There are 24
TCP connections between each (service,machine) pair.

Service A initiates 200kB \blue{transfers using RPCs such that \emph{its} total ingress offered} load is 14\% of
the receivers' rackswitch capacity.  Service B uses 1MB RPCs such that
the \emph{total ingress offered load} on the rackswitch is 15\%, 50\%, 70\% and
$>$100\%.  Table~\ref{tab:latency} shows the 99th percentile
FCT
of 200kB RPCs, while we varied the total load at the rackswitch.  The
mean RPC inter-arrival time $t_{\mu}$ is chosen to match the load, and
the inter-arrivals are sampled uniformly at random between 0 and
$2t_{\mu}$.  The 99th percentile FCT is computed after running for 20
minutes.  \ignore{Note that the total utilization at the rackswitch is limited
to 80\% of its capacity.}

The results in Table~\ref{tab:latency} show, not surprisingly,
that \sys does not significantly affect latency at offered loads
below 100\%.
The table also shows that service A's latency increases with the total
offered load, even though A's offered load is constant, since it
is sharing contention points with service B.  Therefore, one
must assume that B is fully utilizing its maximum bandwidth
allocation when deriving a tail-latency bound for A.

These results also demonstrate that without \sys, the tail
latency of service A can become very large when B tries to
increase its load past its allocated capacity, but \sys protects A's
tail latency.  Naturally, B's own tail latency becomes unbounded when
its load exceeds its guaranteed bandwidth.
Since the over-subscription is at the receiving rackswitch, and the load is
uniformly spread across machines under that switch, \sys's protection is
primarily due to the rack broker's allocations.

\begin{table}[t]\centering
\setlength{\tabcolsep}{2pt}
\begin{tabular}{c|r|r|r|r}
 & \multicolumn{4}{c}{Total offered load at receiver rackswitch} \\
\cline{2-5}
Service &  15\%    & 50\%       & 70\% & $>100$\%   \\\hline
& \multicolumn{4}{c}{\em Without guarantees} \\\hline
A & 2.34ms & 3.50ms & 6.17ms &       \red{\bf 939.79ms}              \\\hline
B & 7.89ms & 7.64ms & 14.30ms &      -               \\\hline\hline
& \multicolumn{4}{c}{\em With EyeQ guarantees} \\\hline
A & 2.33ms & 3.44ms & 6.03ms  &   \red{\bf 701.26ms} \\\hline
B & 9.12ms & 7.71ms & 14.19ms    &  -  \\\hline\hline
& \multicolumn{4}{c}{\em With \sys guarantees} \\\hline
A & 2.34ms & 3.42ms & 6.08ms &     38.74ms$^*$     \\\hline
B & 9.34ms & 7.62ms & 15.07ms &    -       \\\hline
& \multicolumn{4}{c}{\em Bounds (equation \ref{eqn:result})} \\\hline
A & 9.01ms & 15.32ms & 25.53ms & 38.30ms \\\hline
B & 9.77ms & 16.60ms & 27.67ms &  - \\\hline
\end{tabular}
\caption{{The 99th percentile RPC completion time, as a
    function of the total load offered at the receiver rackswitch,
    with and without (w/o) \sys.  The bounds on tail latency are
    computed using equation~\ref{eqn:result}.  As indicated in
    boldface, the latency without \sys suffers at high utilization.}}\myvspace{-1.5em}
\label{tab:latency}
\end{table}

The total load at the rackswitch was limited to 60Gb/s (or about 80\%
of its capacity).  For this experiment, the machine shaper iterates
every 500$\mu$s to recompute $R(t)$, and it converges (in the worst
case) to within 30 iterations, to allocations within 0.01\% of the
ideal rate, regardless of the number of
flows~\cite{jeyakumar2013eyeq}.  In practice, we found that fewer than 15
 iterations suffice.  Thus, the burst size $\sigma$ can be bounded and
is dominated by the convergence time of the congestion control
algorithm.  Table~\ref{tab:latency} also shows the bounds on the flow
completion time computed using Equation~\ref{eqn:result}.  As we can
see, the bounds are useful indicators of performance at high load.
The bounds are a bit pessimistic at lower loads as the bounds are
calculated making no assumptions on the arrival process; with more
assumptions, they can be improved.



\highlight{
\sys improves over EyeQ because it can control tail latency even
when services A and B are given work-conserving allocations, and
therefore could congest the network fabric.  In this scenario,
these services will saturate the rackswitch uplink, causing
overload at congestion point~(iii) in Figure~\ref{fig:contention}.
Since EyeQ assumes a congestion-free core it cannot control tail
latency in such scenarios; instead, EyeQ will default to a TCP-like
behavior as highlighted in Table~\ref{tab:latency}.

In EyeQ, the only solution that would control tail latency would
be to do admission control on service instances, ensuring that
the rack uplink bandwidth is never over-committed.  In other words,
EyeQ could not support work-conserving allocations, and instead
would have to cap (for example) each of the ten instances of
both services to 3Gb/s, so as to limit the total allocation to
60Gb/s and prevent any congestion.  Such static allocations, however,
could waste a large fraction of the scarce uplink bandwidth if either
service is inactive or lightly loaded, while the other is highly
loaded.  In contrast to EyeQ, \sys dynamically adjusts each
service's allocation, thus maximizing resource utilization
under the constraint of limited tail latency.
}
\ignore{
\blue{{\bf Remark}:  The experiment values in Table~\ref{tab:latency} were
the average of 3 runs.  In all but one (marked *) non-highlighted cases, the standard deviation was less than
0.3ms.  Whenever the latency shot up to more than 700ms, the standard deviation was more than 100ms.
However, in one case (\sys $>100$\% load, marked *)
service~A's 99th percentile latencies were 16.07ms, 15.54ms, and 84.62ms.
On further investigation, we found that in the third run, two rack brokers
in the receiving rack had crashed.  However, the remaining eight rack brokers
continued to function and under-estimated the total ingress utilization, which
resulted in a higher latency.}}

\blue{{\bf Remark}: Table~\ref{tab:latency} shows results averaged over 3 runs.
The standard deviation was under 0.3ms, except (a) in the highlighted cases, where the
means exceed 700ms, the std.\ dev.\ is above 100ms, and (b) in the
case marked `*,' where during one run, two (out of ten) rack brokers in the
receiving rack crashed.  In the latter case, the remaining eight
rack brokers continued to function and under-estimated
the total ingress utilization, which resulted in excessive
99th percentile latency.\footnote{The 99th percentile latency from
the three runs were 15.54ms, 16.07ms, and 84.62ms.}  We have not yet
re-run this experiment with 3 bug-free runs.}

\section{Discussion}\label{sec:lessons}
{\bf Timescales for traffic shaping}:
As the authors of EyeQ noted~\cite[\S2.2]{jeyakumar2013eyeq}, rate
control must be fast enough to react before queues build up and increase
latency for competing flows.  However, we also found that such fast
and accurate rate control can also hurt the completion times of RPC
transfers.
This insight is not specific to \sys, but applies to any
bandwidth management scheme targeted at an environment
with relatively high bandwidth and low latency.
\techreport{
\begin{figure}[t]
\centering
\includegraphics[width=0.4\textwidth]{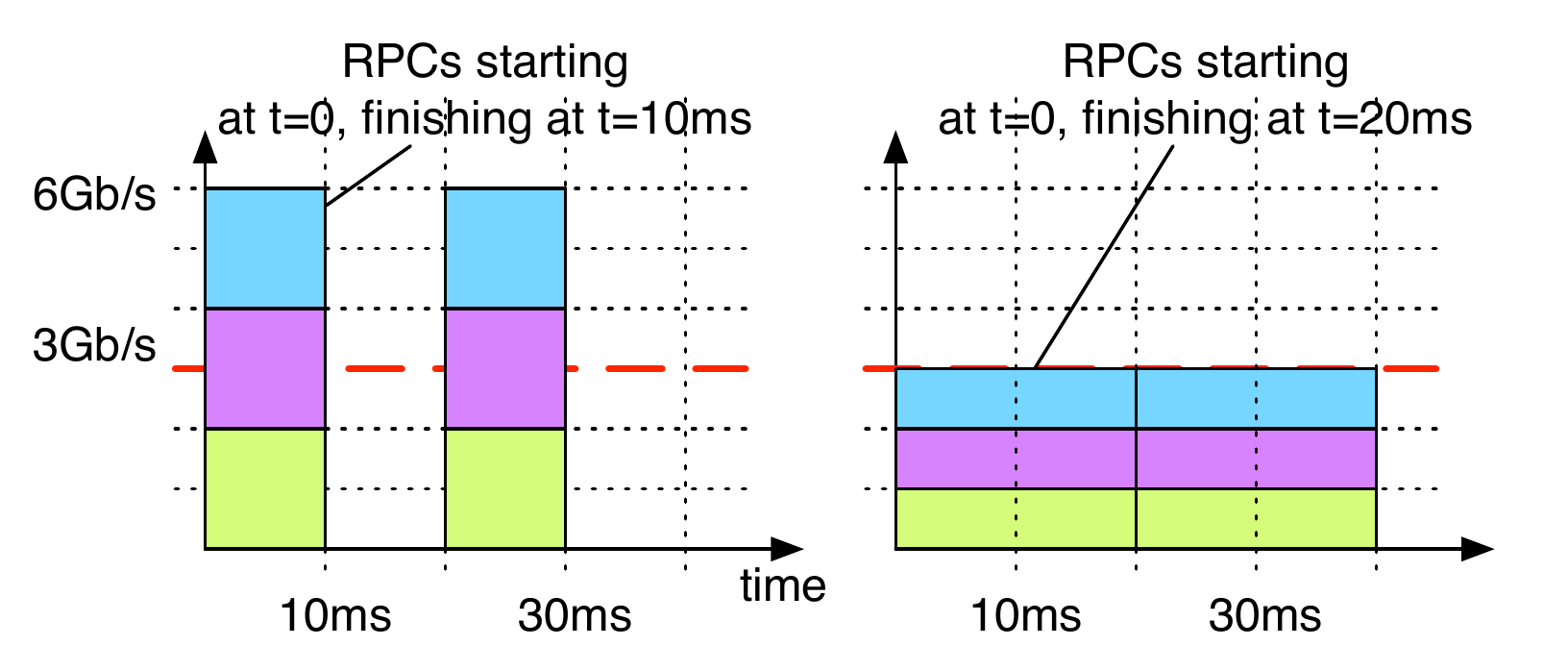}
\caption{{To rate limit or not: Rate limiting and
  accurate pacing improves sharing but increases RPC completion
  times.}}\label{fig:limit-nolimit}\myvspace{-1.5em}
\end{figure}

To understand this, consider the scenario in
Figure~\ref{fig:limit-nolimit}, where service receives three 2.5MB RPCs
at the same time, and then waits for 10ms; thus, its demand
oscillates between 0 and 6Gb/s (measured over 10ms)
every 20ms.
Suppose our policy limits the service to 3Gb/s; over a period of
20ms, its usage complies with the policy.  However, over a 10ms
interval (e.g., from 0--10ms), the service exceeds its
allocation by 3Gb/s.

If we use a fast-acting rate-control mechanism,
the service is always precisely rate limited,
so demand from 0--10ms spills over to the next
interval.  This means that all RPCs take 20ms to finish.
If, however, the service is not rate-limited at all, all of
its RPCs will finish in 10ms, thus halving the maximum
completion time.
This highlights a fundamental
tradeoff between accurate rate guarantees and RPC completion times.
However, it is difficult to upper bound the 
rate-limiting burst size,
as the example in Figure~\ref{fig:limit-nolimit}
still holds if RPC sizes (and timescales) are scaled up by a factor of
10.}

Thus, at the machine shaper, the service rate limiters should have a
configurable burst size for coarse-grained rate limiting, and rate
meters should have a configurable time interval to measure ingress
utilizations.  In
general, the burst size of a latency-sensitive service should be
higher than a collocated throughput-sensitive service.  This burst
size approach is a special case of the more general \emph{fair service
  curves}~\cite{stoica1997hierarchical},
an approach to decouple rate guarantees from delay guarantees.  A rule
of thumb is to set burst sizes larger than the size of RPCs that have
low-latency requirements.  \nontechreport{\blue{We discuss this in detail in our technical report~\cite{techreport}.}}

{\bf Tighter latency bounds}: The latency bound shown in
Equation~\ref{eqn:result} depends both on the capacity $C$ and on the
total load $\rho$ on the network queues, and increases steeply as load
increases (i.e. as $\rho\rightarrow 1$).  Thus, if the resulting
latency bounds are unacceptable, either (a) low-latency RPCs must not
be exposed to load offered by other services, which can be achieved by
prioritization, or (b) network capacity should be
increased.  \sys helps in two ways: (i) it can control the peak
utilization at the rack uplinks/downlinks regardless of the number of
service endpoints under the rack; (ii) \sys serves as a useful
monitoring and protection mechanism for buggy services that might
consume more bandwidth than what is specified according to a policy.
\ignore{
\remark{I don't understand this; how does \sys provide any more
  protection than simply rate-limiting these buggy services according
  to their policy's max limit?  Vimal: rate limiting (at the rack
  level) is the important point here.  By limiting bandwidth
  consumption at the rack, you can use Eqn 2 to bound the tail
  latency.}}





\section{Conclusion}
\label{sec:Conclusion}
We presented \sys, a framework to flexibly and predictably share
datacenter bandwidth across services at the machine, rack, and
fabric levels.  We showed showed how to systematically decompose a sharing
objective into rack-local and machine-local objectives.  On a cluster
of 90 machines, we demonstrated that \sys is able to meet its goals.
We also outlined how predictable bandwidth shares can
help services maintain low tail latencies,
despite potentially malicious traffic patterns of collocated jobs.

{\setlength{\bibsep}{0pt plus 0.5ex}
\ifnatbibstyle
\bibliographystyle{unsrt}
\setlength{\itemsep}{0pt}
\begin{small}
\bibliography{local}
\end{small}
\else
{
\techreport{\renewcommand*{\bibfont}{\small}\printbibliography}
\nontechreport{\printbibliography}
}

\end{document}